\documentclass[11pt]{article}
\usepackage[letterpaper,margin=1in]{geometry}
\usepackage[T1]{fontenc}
\usepackage{lmodern,microtype}
\usepackage{amsmath,amssymb,amsthm,mathtools}
\usepackage{booktabs,enumitem,xcolor,pgfplots,hyperref}
\pgfplotsset{compat=1.18}
\hypersetup{colorlinks=true,linkcolor=blue,citecolor=blue,urlcolor=blue,
  pdftitle={Online Edge-Weighted Matching Beyond 0.55},
  pdfauthor={Alireza Farhadi and Mohammad Hajiaghayi}}
\setlist{itemsep=2pt,topsep=4pt}
\newtheorem{theorem}{Theorem}
\newtheorem{lemma}[theorem]{Lemma}
\newtheorem{proposition}[theorem]{Proposition}
\newtheorem{corollary}[theorem]{Corollary}
\theoremstyle{definition}
\newtheorem{definition}[theorem]{Definition}
\newtheorem{remark}[theorem]{Remark}
\DeclareMathOperator{\OPT}{OPT}
\DeclareMathOperator{\ALG}{ALG}
\DeclareMathOperator{\Cov}{Cov}
\DeclareMathOperator{\Bin}{Bin}
\newcommand{\E}{\mathbb{E}}
\newcommand{\R}{\mathbb R}
\newcommand{\one}{\mathbf 1}
\newcommand{\st}{\mathrm{st}}
\newcommand{\icx}{\mathrm{icx}}
\newcommand{\BestRatio}{0.556195}
\newcounter{algorithm}
\newcounter{pseudoline}
\newcommand{\pc}{\stepcounter{pseudoline}%
  \makebox[1.3em][r]{\scriptsize\arabic{pseudoline}}}
\newenvironment{pseudocode}[2]{%
  \begin{figure}[!tb]\small
  \refstepcounter{algorithm}\label{#2}%
  \hrule\vspace{5pt}\noindent
  \textbf{Algorithm \thealgorithm.\enspace #1}\par
  \vspace{2pt}\setcounter{pseudoline}{0}%
  \begin{tabbing}
  \hspace{1.8em}\=\hspace{1.3em}\=\hspace{1.3em}\=
  \hspace{1.3em}\=\hspace{1.3em}\=\hspace{1.3em}\=\kill
}{%
  \end{tabbing}\vspace{-3pt}\hrule\end{figure}%
}
\title{Online Edge-Weighted Matching Beyond $0.55$}
\author{Alireza Farhadi\thanks{Profound.}
\and Mohammad Hajiaghayi\thanks{University of Maryland, College Park.
Email: \texttt{hajiagha@umd.edu}.}}
\date{}
\begin{document}
\maketitle
\begin{abstract}
We give a $\BestRatio$-competitive randomized algorithm for edge-weighted
online bipartite matching with free disposal, improving the $0.5368$
guarantee of Blanc and Charikar. The result holds for arbitrary
nonnegative edge weights and any input sequence fixed independently
of the algorithm's randomness. Our algorithm builds on online correlated
selection, introduced by Fahrbach, Huang, Tao, and Zadimoghaddam,
and its multiway extension due to Blanc and Charikar.

Our main contribution is a state-dependent primal--dual analysis of
correlated selection. At each weight threshold, we track the state
of the selection process jointly with an auxiliary failure event.
The analysis accounts for arrivals below the threshold, which change
the state without improving the matching value at that threshold.
A potential function bounds the dual increase in both cases.
We instantiate the analysis with an explicit win map that assigns
a common score to an initial range of states. Its selection
inequalities are verified by a finite
computer-assisted proof covering every required candidate multiplicity
and every real threshold; the verifiers are available in an accompanying
repository.
Triangular unit-weight inputs asymptotically attain the guarantee
of the underlying ideal algorithm, proving that its analysis is tight.
\end{abstract}

\section{Introduction}
\label{sec:introduction}

Online bipartite matching asks how much of an optimal allocation can
be recovered when one side of the graph is revealed over time.
For unweighted matching, the Ranking algorithm of Karp, Vazirani,
and Vazirani~\cite{KVV90} achieves the optimal competitive ratio
$1-1/e$. The same ratio is attainable when weights belong to the
offline vertices~\cite{AGKM11}. With arbitrary edge weights and no
free disposal, an early assignment can prevent an arbitrarily more
valuable later assignment. Free disposal allows each offline vertex
to retain its most valuable assigned edge. This is a
natural model for display advertising, where an advertiser may receive
several impressions but pays only for its most valuable assignment
\cite{FKMMP09}.

Even with free disposal, the greedy $1/2$ guarantee resisted improvement
for more than a decade. Fahrbach, Huang, Tao, and
Zadimoghaddam~\cite{FHTZ22} broke this barrier with a
$0.5086$-competitive algorithm and introduced online correlated
selection (OCS).
Gao et al.~\cite{GHHNYZ21} obtained a $0.519$ guarantee, and Blanc
and Charikar~\cite{BC21} introduced multiway OCS and obtained $0.5368$.
The OCS viewpoint separates two tasks. A sampling rule chooses the
candidates for the current arrival, and a selection procedure chooses
one integral assignment from those candidates.
Correlating the selections across arrivals can make repeated requests
to a vertex more effective than independent rounding.

For weighted matching, however, a vertex is not simply covered or
uncovered. An edge of weight $w$ can help at every level below $w$
and at none of the levels above it. A useful selection process must
therefore work for many overlapping subsequences of arrivals.
Moreover, arrivals outside one such subsequence still advance the
process. A competitive analysis must account for these changes of
state without charging them to matching gain at that threshold.

We address this issue by keeping the state distribution in the dual
analysis. This removes a restriction in the earlier reduction that
limits the benefit of sampling many candidates.

\subsection{Our results}

The offline side $L$ of a bipartite graph is known in advance.
Online vertices arrive in an arbitrary fixed order and reveal their
incident nonnegative edge weights. Each arrival is assigned to at most
one offline vertex. An offline vertex may receive several assignments;
only its heaviest assigned edge contributes to the objective. The
benchmark $\OPT$ is the weight of a maximum matching in the entire
input graph. The input is fixed independently of the algorithm's
random choices.
Missing edges may be treated as zero-weight edges; an assignment to
one is implemented by discarding the arrival.

\begin{theorem}\label{thm:main}
There is a randomized $\BestRatio$-competitive algorithm for
edge-weighted online bipartite matching with free disposal. For every
finite input graph and every fixed arrival order,
\[
                         \E[\ALG]\ge\BestRatio\,\OPT.
\]
For rational input weights, the algorithm has a finite implementation
with expected polynomial running time, with fixed preprocessing
depending only on the constant in the guarantee.
\end{theorem}

The graph-independent preprocessing is large.
Section~\ref{sec:parameters} gives the constants and their certificates.
Section~\ref{sec:algorithm} gives the finite algorithm, its stored
state, and pseudocode for both sampling and assignment.
The verification programs and Lean sources are available in the
artifact repository.\footnote{\url{https://github.com/farhadi94/online-edge-weighted-matching}}
Appendix~\ref{sec:verifier} gives reproduction instructions.

The proof has two components. First, we give a state-dependent
primal--dual theorem for a family of geometric-age selection processes.
The process records an age: the number of steps since its most recent
reset. A win map $W$ assigns a desired win probability to each output
of this process. Let $T_W$ be the substochastic matrix that retains
only transitions on which an auxiliary selection trial fails, and
let $\pi$ be the stationary age distribution. When $W$ satisfies
the tournament inequalities, the competitive guarantee is
\begin{equation}\label{eq:intro-score}
 \gamma(W)=1-\frac1{m+1}\sum_{n\ge0}
       \left(\frac m{m+1}\right)^n\pi T_W^n\one.
\end{equation}
Here $m$ is the number of candidates sampled on an arrival.
This expression already appears in
Blanc--Charikar's reduction~\cite[Theorem~7]{BC21}, under additional
shape and gap conditions. To see the restriction imposed by the gap
condition, write $F(n)=\pi T_W^n\one$. For a map with stationary
mean $1/m$, we have $F(1)-F(0)=-1/m$. When $\gamma>1/2$, the gap
condition in~\cite[Equation~(23)]{BC21} requires
\[
 m\le\frac{1-\gamma}{2\gamma-1},\qquad\text{or equivalently}\qquad
 \gamma\le\frac{m+1}{2m+1}.
\]
This upper bound tends to $1/2$ as the number of candidates grows.
Theorem~\ref{thm:direct-age} establishes the same
guarantee for the geometric-age family without this restriction
or the scalar shape assumptions. Its invariant accounts for
intervening arrivals through their effect on the state distribution.
Theorem~\ref{thm:ideal-triangular-tightness} shows that, for
maps of stationary mean $1/m$ with positive baseline, $\gamma(W)$ is the exact
worst-case ratio of the specified ideal algorithm. A deterministic
family of triangular unit-weight graphs attains it asymptotically.

Second, we choose the win map. For fixed $m$ and reset probability,
Theorem~\ref{thm:paper-pooling} characterizes an optimal nondecreasing
map under the necessary calibration constraint for a vertex sampled
once. It assigns the same baseline score to an initial segment of ages,
modifies at most one boundary age, and preserves the remaining
quantile-bin scores. The map is explicit once the baseline is fixed.
Proposition~\ref{prop:regenerative-purification} extends this optimization
to a broader class of processes with independent resets by averaging
over their internal state conditional on the time since the last reset.

A vertex sampled more than once imposes further selection constraints;
the optimization theorem alone does not establish these. We verify
them for the map used in Theorem~\ref{thm:main}.
Section~\ref{sec:parameters} fixes the constants,
Appendix~\ref{sec:certificate} proves the finite certificate's
sufficiency, and Appendix~\ref{sec:verifier} identifies the repository
verifiers.

\subsection{Related work}

\paragraph{Online matching and free disposal.}
The OCS literature has developed several ways to correlate online
choices while preserving feasible assignments. Gao, He, Huang, Nie, Yuan, and Zhong~\cite{GHHNYZ21} strengthened
two-way OCS and obtained a $0.519$ guarantee.  Shin and An~\cite{SA21}
constructed a three-way OCS, obtaining $0.5093$ directly and $0.5132$
when combined with the improved two-way OCS.  Blanc and
Charikar~\cite{BC21} developed multiway OCS and obtained a
$0.5368$-competitive algorithm.  Their reduction also showed that the
original two-way formulation cannot give a ratio above $0.5239$.
For complete bipartite graphs with product-form weights
$w_{ij}=a_i b_j$, Charikar, Henzinger, and Nguyen~\cite{CHN14}
obtained a $0.5664$-competitive algorithm. That result exploits the
decomposable weight structure; the theorem here permits arbitrary
edge weights on arbitrary bipartite graphs.
The surveys of Mehta~\cite{Mehta13} and Huang, Tang, and
Wajc~\cite{HTW24} give broader background on the models and techniques
of online matching and ad allocation.

\paragraph{Primal--dual analysis and submodular allocation.}
Devanur, Jain, and Kleinberg~\cite{DJK13} gave a randomized
primal--dual analysis of Ranking and its vertex-weighted extension,
with dual feasibility in expectation.  For free disposal, Devanur,
Huang, Korula, Mirrokni, and Yan~\cite{DHKMY16} developed an analysis
using a level function.  The threshold decomposition in the present
paper belongs to this line of work and to the complementary
cumulative-distribution viewpoint used by Fahrbach et
al.~\cite{FHTZ22}.  Our dual credit additionally records the
joint distribution of the selection state and the event of auxiliary
failure. Fractional free-disposal algorithms admit the stronger
$1-1/e$ guarantee. Recent treatments include the primal--dual
frameworks surveyed by Niazadeh and Udwani~\cite{NU26} and the
minimax framework of Kesselheim, Molinaro, Patton, and
Singla~\cite{KMPS26}. These fractional guarantees do not supply the
integral rounding required here.

Free-disposal matching is a special case of online submodular welfare
maximization: the valuation of an offline vertex for its assigned set
is the maximum incident edge weight in that set.  The special form
of these valuations matters.  Kapralov, Post, and
Vondr\'ak~\cite{KPV13} rule out polynomial-time competitive ratios
strictly above $1/2$ for general coverage valuations, unless
$\mathrm{NP}=\mathrm{RP}$. Udwani and Wajc~\cite{UW26} recently
proved an unconditional $1/2$ upper bound for general online submodular
welfare in the standard model, where value queries involve already
arrived items. Our improvement uses the maximum-weight valuation of
matching; it does not extend to arbitrary monotone submodular valuations.

\section{Overview of the Proof}
\label{sec:techniques}

\subsection{A state-dependent analysis for general edge weights}

\paragraph{From weights to coverage.}
For an offline vertex $i$, let $M_i$ be the largest weight assigned
to it. Its expected contribution is
\[
                  \E[M_i]=\int_0^\infty\Pr(M_i\ge w)\,dw.
\]
Thus we can study, simultaneously for every $w$, whether an arrival
of weight at least $w$ has selected $i$. This complementary cumulative
distribution function viewpoint is standard in weighted online
matching~\cite{DHKMY16,FHTZ22}. It explains why negative correlation
is useful: repeated opportunities above a threshold should be less
likely to fail together. It also identifies the main obstacle.
An arrival below the threshold does not cover it, but it can change
the distribution of the next selection opportunity.

\paragraph{The selection process.}
Give each offline vertex an independent process whose state is its age
since the last reset. Each occurrence of the vertex among the sampled
candidates advances the process by one step, called an exposure.
The age increments with probability $q=1-p$ and resets to zero with
probability $p$. A reset at age $z$ produces a score
$W(z)$, representing a desired win probability; a nonreset
produces the baseline $b=W(-1)$.
Write $X_k$ for the desired win probability at step $k$. We take $W$
to be nondecreasing: a longer wait before a reset gives a larger score,
while a reset returns the process to age zero. This dependence between
successive scores is what we use to improve on independent selection.

On each arrival, the algorithm samples $m$ candidates independently
from a probability vector $v$. If vertex $i$ appears $r$ times, its
process takes $r$ consecutive steps. Its combined desired win probability is
$1-\prod_{k=1}^r(1-X_k)$.
A monotone tournament of Blanc and Charikar~\cite{BC21} converts the
combined probabilities into one actual winner. The vector $v$ is a deterministic
function of the input prefix and the averaged states. It is independent
of all realized candidate multisets, seed states, and tournament
outcomes. Consequently, after conditioning on the candidate multisets,
the selection processes retain the dependence properties needed in
the analysis.

The tournament bounds the probability that a vertex remains unselected
on any fixed set of arrivals by the expected product of its auxiliary
failure probabilities. In particular, the bound applies to the arrivals
whose edge to that vertex has weight at least a chosen threshold.

\paragraph{The residual potential.}
Fix a vertex $i$ and a threshold $w$. Let $\nu(z)$ be the probability that
all auxiliary trials for edges of weight at least $w$ have failed and the current
age is $z$. This is a subprobability distribution: its total mass is
the current auxiliary failure probability. Write $P$ for the age
transition matrix and $T$ for the substochastic matrix that also
requires the auxiliary trial to fail. On an arrival with edge weight
$w_i$, each step sends $\nu$ to $\nu T$ if $w\le w_i$, and to
$\nu P$ if $w>w_i$.

The state $\nu$ records more than its total mass. Two thresholds can
have the same failure probability but different age distributions.
To account for this difference, continue the auxiliary trials for an
independent geometric number of steps with mean $m$. Let $h(z)$ be
the probability of at least one success, starting at age $z$.
Starting at a larger age gives at least as large a success probability
at the first reset; after that reset, the processes can be coupled to
agree. Thus $h$ is nondecreasing.

Let $\pi$ be the stationary age distribution and put $\gamma=\pi h$.
The residual potential of the threshold is $\nu h$: the expected
success probability that remains on its uncovered mass. We assign
the complementary quantity $\gamma-\nu h$ as offline dual credit.
It starts at zero because initially $\nu=\pi$. Solving the one-step
recurrence for the independent stopping experiment gives
\[
                  h=\one-\bigl(I-m(T-I)\bigr)^{-1}\one.
\]
Proposition~\ref{prop:potential-optimality} shows that this choice
maximizes $\pi h$ among bounded linear potentials satisfying the local
update inequality for $w\le w_i$ on states satisfying the invariant below.

\paragraph{Credit monotonicity.}
The proof uses the complementary vector $\theta=\pi-\nu$.
We show that every reachable $\theta$ is a nonnegative linear combination of
$\pi$ and the distributions $\delta_0P^n$, $n\ge0$. The latter is the
distribution of a geometric age truncated at $n$; it becomes
stochastically older when another ordinary exposure is taken.
This representation is preserved by both updates.
A $P$-update advances each component by one age transition.
A $T$-update also contributes a
stationary component from the baseline success probability and a
fresh reset component from the additional success probability at a
reset. The existing components are rescaled and advanced; all
coefficients remain nonnegative. Since each component becomes
stochastically older under an age transition, the representation gives
$\theta Ph\ge\theta h$, equivalently $\nu Ph\le\nu h$.
Since $0\le T\le P$ and $h\ge0$, a $T$-update also
satisfies $\nu Th\le\nu Ph\le\nu h$. Thus both updates increase
the credit $\gamma-\nu h$. This invariant handles arbitrary
interleavings of arrivals above and below the threshold.

\paragraph{The two update inequalities.}
Suppose the current vertex is sampled with probability $x$.
Its number of exposures is $\Bin(m,x)$. If $w\le w_i$, the
updated state is $\nu(I+x(T-I))^m$. Write $y(x)$ for its mass and
$a(x)$ for its residual potential. The recurrence defining $h$ gives
\[
                     m(T-I)(h-\one)=h.
\]
After differentiating the binomial update and using the invariant,
this identity yields
\[
                         -y'(x)+a'(x)\ge a(x).
\]
The decrease in auxiliary failure probability thus pays for the
increase in offline credit and leaves a contribution toward the
online vertex's dual variable.

If $w>w_i$, set
$a(x)=\nu(I+x(P-I))^m h$. There is no coverage gain, although the
age distribution still changes. The same recurrence shows
that $k=(I-m(P-I))h$ is nonnegative and has stationary expectation
$\gamma$. Every state $\nu$ reached by the algorithm satisfies
$0\le\nu\le\pi$. Since $k\ge0$, we have $\nu k\le\pi k=\gamma$.
Together with credit monotonicity, this gives
\[
                         -a'(x)\le\gamma-a(x).
\]
After integration, the credit created when $w>w_i$ is
at most the sampling probability times its post-arrival credit. This
term is subtracted in the sampling deficit below. The accounting
therefore includes state changes above the current edge weight,
where no matching gain is available.

\paragraph{Water filling.}
Integrate the threshold credits to obtain one offline dual variable
$\alpha_i$ per vertex, and integrate one minus the auxiliary miss
probability to obtain its contribution $\mathcal P_i$ to the surrogate
primal objective. This objective lower-bounds the expected matching
value. For an edge of weight $w_i$ and sampling probability $x$, the
two update inequalities combine
into
\[
 \Delta\mathcal P_i-\Delta\alpha_i
       \ge x\bigl(\gamma w_i-\alpha_i[x]\bigr),
\]
where $\alpha_i[x]$ is the credit after the hypothetical update.
This suggests the terminal deficit
$d_i(x)=\gamma w_i-\alpha_i[x]$. These deficits are continuous
and nonincreasing. A large deficit means that the edge's dual
constraint still requires a large contribution from the arriving
vertex. The algorithm increases the coordinates with the
largest deficit until their total is one; a zero-weight dummy absorbs
any remaining probability. Every used coordinate has the same terminal
deficit $\beta$. This gives dual feasibility, while integrating the
two update inequalities pays for $\sum_i\Delta\alpha_i+\beta$.
Summing over arrivals proves the competitive guarantee
$\gamma$.

\paragraph{Tournament feasibility.}
The remaining condition belongs to the selection process. For each
$1\le r<m$, let $V_r=\prod_{k=1}^r(1-X_k)$. The tournament exists
when
\[
 \E[(s-V_r)_+]\le
 H_r(s):=s-\frac{m-r}{m}
       +\frac{m-r}{m}(1-s)^{m/(m-r)}
 \quad(0\le s\le1).
\]
The singleton condition follows from the quantile construction.
For repeated candidates, condition on the first reset. The remainder
starts at age zero and has bounded age. Its minimum and mean failure
products satisfy short recurrences. Replacing this fresh product by
a two-point distribution with the same mean, supported at its lower
and upper bounds, increases every convex positive-part expectation.
This avoids enumerating the reset patterns.

The initial age is unbounded. Conditional Jensen combines its older
quantile bins into an explicit integral, while the pooled ages remain
one point mass. Conditioning on the first reset thus gives a convex upper
bound with at most $2r$ age-averaged terms and a point mass for the
no-reset case. A quadratic
estimate handles thresholds near zero. Elsewhere, chords between
certified endpoint values bound the curve. Minimizing the gap between
each chord and $H_r$ reduces its validity to an exact integer-power
inequality. A finite rational partition therefore covers every real
threshold; Appendix~\ref{sec:certificate} gives the construction.

\subsection{Choosing the win map}
\label{sec:pooling-intuition}

For fixed $m$ and $p$, the necessary singleton constraint is
$W(Y)\preceq_{\icx}U^{m-1}$, where $Y$ is an output of the stationary reset process,
$U$ is uniform on $[0,1]$, and $\preceq_{\icx}$ denotes increasing
convex order. To see why pooling arises, fix a baseline
$b\in[q^{m-1}/m,1/m]$ and consider maps of mean $1/m$. Their integrated lower-quantile curve
$S(u)$ must satisfy $S(u)\ge bu$ and $S(u)\ge u^m/m$.
The smallest permitted values at the seed's quantile-bin endpoints
are therefore given by
\[
                         \phi_b(u)=\max\{bu,u^m/m\}.
\]
The successive secant slopes define a nondecreasing win map. The
linear part assigns a common score to the youngest ages, and the
remaining slopes preserve the older quantile-bin scores. A renewal
formula shows that, at fixed baseline, $\gamma(W)$ is nondecreasing
in the stationary upper-tail score sums. The pooled map
maximizes these sums at fixed baseline. Theorem~\ref{thm:paper-pooling}
proves this comparison, including maps whose mean is less than $1/m$,
and reduces the optimization to the choice of $b$.

Age also suffices for a broader class of processes with independent resets. Conditional
on the reset times, each cycle contributes at most one nonconstant
factor to a failure product. Replacing its terminal score by its
conditional mean preserves avoidance probabilities and, by Jensen's
inequality, preserves the tournament constraints.
Proposition~\ref{prop:regenerative-purification} proves this reduction
by averaging over the internal randomness between resets.
The subsequent pooled replacement optimizes only the singleton
relaxation; its higher-multiplicity constraints still require the
certificate described above.

\section{Correlated Selection and Weight Thresholds}\label{sec:selection}

\subsection{The stationary geometric-age seed}

Fix $m\ge2$, a reset probability $p\in(0,1)$, and $q=1-p$.
Start $Z_1$ with distribution $\pi(z)=pq^z$, $z\ge0$.
At exposure $k$, independently draw a reset bit $B_k$ with probability
$p$ of being one. If $B_k=1$, output $Y_k=Z_k$ and set
$Z_{k+1}=0$. Otherwise output $Y_k=-1$ and set $Z_{k+1}=Z_k+1$.
Every offline vertex gets an independent seed.

Let $W(-1)=b\in[0,1)$ and let $W(z)$ be nondecreasing in $z\ge0$,
with $b\le W(z)\le1$. The desired win probability at exposure $k$ is
$X_k=W(Y_k)$. The raw sequence $(Y_k)$ is stationary and negatively
associated~\cite[Section~10.1]{BC21};
Lemma~\ref{lem:paper-age-na} gives a short independent proof.
Coordinatewise increasing maps preserve negative association, so the
same holds for $(X_k)$.

\begin{definition}[Negative association]
A finite family $(X_k)_{k\in I}$ is negatively associated if, for
disjoint index sets $A,B\subseteq I$ and increasing functions $g,h$
for which the expectations exist,
\[
 \Cov\bigl(g((X_k)_{k\in A}),h((X_k)_{k\in B})\bigr)\le0.
\]
An infinite sequence is negatively associated if all its finite
subfamilies are.
\end{definition}

Negative association is a property of the stationary win process,
not an assertion that the algorithm's choices remain negatively
associated after conditioning on their past outcomes. The analysis
never requires such a conditional assertion.

\subsection{The monotone tournament}

For random variables $A,B$ in $[0,1]$, we write
$A\preceq_{\icx}B$ when $\E[\varphi(A)]\le\E[\varphi(B)]$
for every increasing convex function $\varphi$. Equivalently,
$\E[(A-t)_+]\le\E[(B-t)_+]$ for every $t\in[0,1]$.
The positive-part function measures the expected excess above
a given level.

For $r=1,\ldots,m-1$, put
\[
 V_r=\prod_{k=1}^r(1-X_k),\qquad U_r=1-V_r.
\]
The following sufficient condition is exactly the one used in
Blanc--Charikar's tournament construction.

\begin{lemma}[Tournament criterion~\cite{BC21}]
\label{lem:tournament}
Suppose that, for every $r=1,\ldots,m-1$ and every $t\in[0,1]$,
\begin{equation}\label{eq:tournament-original}
 \E[(U_r-t)_+]
 \le 1-t-\frac{m-r}{m}\left(1-t^{m/(m-r)}\right).
\end{equation}
Then an $m$-candidate multiset admits a randomized tournament with
the following properties. If vertex $i$ has multiplicity $r_i$ and
its next $r_i$ scores are $X_{i,1},\ldots,X_{i,r_i}$, its combined
desired win probability is $Q_i=1-\prod_{k=1}^{r_i}(1-X_{i,k})$.
Conditional only on $Q_i=u$, its
probability of winning is at least $u$, where the probability
averages over independent stationary opponent blocks and fresh
tournament randomness. For a fixed tuple of desired win probabilities, the
tournament kernel gives each vertex a winning probability that is
nondecreasing in its own desired win probability and nonincreasing in every
opponent's desired win probability. A vertex
of multiplicity $m$ wins deterministically.
\end{lemma}

The reference distribution for multiplicity $r$ has cumulative distribution function
$t^{r/(m-r)}$. Raising a reference draw
to the power $1/(m-r)$ produces the maximum of $r$ independent
uniform draws. Comparing these strengths across a partition of $m$
gives the desired conditional winning probability exactly. Inequality
\eqref{eq:tournament-original} says that the proposed desired win probability is
below the reference draw in increasing convex order. The monotone
submartingale-coupling theorem of M\"uller and
R\"uschendorf~\cite[Theorem~4.5(a)]{MR01}, used
in~\cite[Theorem~8 and Lemma~9.13]{BC21}, then transfers the reference strengths to the
proposed distribution. We use this coupling theorem in its published form.

\subsection{A nonselection bound for arbitrary subsets of rounds}

Given a deterministic probability vector $v=(v_i)$, sample $m$
candidates independently from the categorical distribution $v$,
using fresh randomness independent of all seeds and earlier draws.
The multiplicity of $i$ is $K_i\sim\Bin(m,v_i)$. Consume that many
consecutive exposures from $i$'s seed and run the tournament.
The chosen vertex receives the arriving online vertex.

The algorithm below chooses every probability vector from
deterministic states that average over all earlier multiplicities.
For a fixed input, the entire vector sequence is therefore
independent of the realized multisets and seed outcomes. Fix
this sequence and condition on all the sampled multisets. For any fixed vertex $i$
and any deterministic set $S$ of rounds, condition additionally on $i$'s
desired win probabilities in those rounds. The opponents' seed sequences remain
independent of $i$'s sequence. They are jointly negatively associated,
and losing to them in disjoint rounds is increasing in disjoint sets
of their seed coordinates. With independent tournament randomization
on different rounds, the conditional probability of losing every
round in $S$ is at most the product of the individual loss
probabilities. Tournament consistency gives
\begin{equation}\label{eq:arbitrary-subset-proxy}
 \Pr[i\text{ never selected in }S]
 \le \E\left[\prod_{k\in E(S)}(1-X_k)\right],
\end{equation}
where $E(S)$ is the union of the seed exposures consumed in those
rounds. This is the arbitrary-subset assertion of
\cite[Lemma~9.11]{BC21}, followed by averaging over the candidate
multisets. Repeated occurrences consume distinct exposure indices.

For a weight threshold $w$, take
$S=\{j:w_{ij}\ge w\}$. Auxiliary independent success coins with
probabilities $X_k$, used only for the analysis, turn the right side
of~\eqref{eq:arbitrary-subset-proxy} into a miss probability. The
reduction below tracks that probability jointly with the age. The
auxiliary failure probability upper-bounds the actual nonselection
probability; equality is not required.

\section{The Algorithm}
\label{sec:algorithm}

On each arrival, deterministic accounting vectors define a sampling
distribution and are updated by averaging over its candidate counts.
The algorithm then samples $m$ candidates, advances their sampled
seeds, and selects one winner. The sampled outcomes do not feed back
into the accounting vectors. This separation preserves the
arbitrary-subset guarantee from Section~\ref{sec:selection}.

The description is parameterized by $m,p,W$, an age cap $C$, a score
scaling factor $\eta$, and a water-filling accuracy $\epsilon$.
Section~\ref{sec:parameters} sets these constants. We use the finite
capped and scaled process directly; the subsequent analysis first
establishes the ideal guarantee and then bounds the implementation
loss in Appendix~\ref{sec:finite-implementation}.

\subsection{What is precomputed}
\label{sec:algorithm-preprocessing}

Write $q=1-p$, $b_f=\eta W(-1)$, and $W_f(z)=\eta W(z)$ for
$0\le z\le C$, and put $z^+=\min\{z+1,C\}$. On $\{0,\ldots,C\}$,
the transition matrix $P_f$ and substochastic failure matrix $T_f$ satisfy
\[
 \begin{array}{ll}
 P_f(z,z^+)=q,&P_f(z,0)=p,\\
 T_f(z,z^+)=q(1-b_f),&T_f(z,0)=p(1-W_f(z)).
 \end{array}
\]
The stationary distribution is $\pi_f(z)=pq^z$ for $z<C$ and
$\pi_f(C)=q^C$. Preprocessing computes the exact rational quantities
\begin{equation}\label{eq:algorithm-finite-potential}
 h_f=\one-\bigl((m+1)\mathrm{Id}-mT_f\bigr)^{-1}\one,
 \qquad \Gamma=\pi_f h_f.
\end{equation}
Here $\mathrm{Id}$ is the identity matrix on the finite age space.
These are the finite versions of the potential and guarantee in
Section~\ref{sec:techniques}. The sampling rule uses this exact
$\Gamma$; the decimal constant in Theorem~\ref{thm:main} is a lower
bound on the final guarantee, not a replacement for it. A backward
recurrence in Appendix~\ref{sec:finite-potential-computation} computes
$h_f$ in $O(C)$ rational arithmetic operations.

The remaining preprocessed objects are finite tournament tables.
For each multiplicity $r<m$ and possible combined desired win probability $u$, a
table specifies a distribution $K_r(u,\cdot)$ over ordered strength
bins. A second table gives the winner probabilities when several
candidates occupy the highest sampled bin. Thus the online tournament
requires only table lookups and draws from rational distributions.
Appendix~\ref{sec:explicit-tournament-preprocessing} gives the
stationary-block dynamic program, the rational coupling LP, and the
polynomial integrals that construct these tables. Their sizes can be
large; all depend only on the fixed seed parameters, not on the graph
or its edge weights.

\subsection{The state stored at an offline vertex}
\label{sec:algorithm-state}

For each real offline vertex $i$, maintain the following objects.
The dummy vertex $\bot$ has only the sampled age.
\begin{itemize}
\item A \emph{sampled age} $Z_i\in\{0,\ldots,C\}$, initialized
  independently from $\pi_f$. It advances only when $i$ occurs among
  the sampled candidates.
\item A \emph{retained edge} $e_i$ and its weight $M_i$, initially
  empty and zero. This record implements free disposal.
\item A collection of \emph{accounting vectors} $\nu_{i,I}$, one for
  each interval $I$ between consecutive positive weights revealed at
  $i$, with zero included as the first endpoint. The vector records
  the auxiliary miss event jointly with the age, averaged over all
  candidate samples and seed outcomes.
\end{itemize}

The accounting vectors are deterministic functions of the input
prefix. They are subprobability vectors, so their entries need not
sum to one and are never normalized. They determine the sampling
distribution. The sampled ages and retained edges are used only to
execute the random assignment and record its value.

Initially no finite intervals are stored. The implicit tail above
the largest weight revealed at $i$ always has vector $\pi_f$ and
contributes zero credit. When an edge of weight $w_i$ arrives, insert
$w_i$ as an endpoint if it is positive and new. An interior insertion
splits one interval into two copies of its old vector. An insertion
above all previous weights creates a new finite interval with vector
$\pi_f$. Denote this operation by $\mathrm{Refine}(i,w_i)$ and the
resulting set of finite intervals by $\mathcal I_i$. At arrival $j$
there are at most $j$ such intervals per vertex.

Every interval now lies entirely on one side of $w_i$. Define
\begin{equation}\label{eq:algorithm-interval-kernel}
 K_{i,I}=
 \begin{cases}
 T_f,&\sup I\le w_i,\\
 P_f,&\inf I\ge w_i,
 \end{cases}
 \qquad R_{i,I}(x)=(1-x)\mathrm{Id}+xK_{i,I}.
\end{equation}
Below the current weight,
an exposure must be an auxiliary miss to remain in the accounting
vector. Above that weight, an exposure simply advances the age. In
particular, the intervals above $w_i$ also receive an update.

\subsection{Choosing the candidate probabilities}
\label{sec:algorithm-probabilities}

Suppose the sampling probability of $i$ were $x$. Its multiplicity
would be $\Bin(m,x)$, so the averaged vector on interval $I$ would
be $\nu_{i,I}R_{i,I}(x)^m$. The resulting total credit is
\begin{equation}\label{eq:algorithm-credit}
 A_i(x)=\sum_{I\in\mathcal I_i}|I|
       \left[\Gamma-\nu_{i,I}R_{i,I}(x)^m h_f\right],
 \qquad d_i(x)=\Gamma w_i-A_i(x).
\end{equation}
Here $|I|$ is the interval length. All hypothetical evaluations use
the same vectors present just after $\mathrm{Refine}$. Trying a
sampling probability does not modify those vectors.

The deficit $d_i(x)$ measures how much of the current edge's dual
constraint remains after giving $i$ probability $x$. It is
nonincreasing in $x$, as proved in Section~\ref{sec:state-duality}.
Water filling repeatedly gives more probability to a vertex with the
largest deficit. We use the finite rational rule in
Algorithm~\ref{alg:probabilities}: at arrival $j$, allocate probability
in increments of $1/N_j$, where
$N_j=\lceil mj(j+1)/\epsilon\rceil$. A dummy coordinate has deficit
identically zero. Fix a deterministic tie order once, with the dummy
first. After $N_j$ increments the probabilities sum to one. The grid
approximately equalizes the terminal deficits. Its total loss in
competitive ratio is at most $\epsilon$, independently of the number
of arrivals.

Each deficit is an explicit polynomial. To evaluate it without taking
a matrix power for every trial value of $x$, first compute
\begin{equation}\label{eq:algorithm-bernstein}
 \begin{split}
 c_{i,k}&=\sum_{I\in\mathcal I_i}|I|
        \left[\Gamma-\nu_{i,I}K_{i,I}^{k}h_f\right]
                         \qquad(0\le k\le m),\\
 A_i(x)&=\sum_{k=0}^m\binom mkx^k(1-x)^{m-k}c_{i,k}.
 \end{split}
\end{equation}
The first line conditions on exactly $k$ exposures; the second
averages over their binomial count. Sparse row-vector products
compute all coefficients. Precomputing the powers of $x$ and $1-x$
up to degree $m$ makes each subsequent evaluation use $O(m)$ rational
arithmetic operations. A max heap stores the current
deficits. An increment changes only the chosen coordinate's key.

\begin{pseudocode}{Choosing the sampling probabilities}{alg:probabilities}
\pc\>\textbf{Input:} arrival $j$, weights $(w_i)$, and frozen interval vectors.\\
\pc\>\textbf{for each} real vertex $i$ \textbf{do}\\
\pc\>\>Set $c_{i,k}\gets0$ for $k=0,\ldots,m$.\\
\pc\>\>\textbf{for each} $I\in\mathcal I_i$ \textbf{do}\\
\pc\>\>\>Set $u\gets\nu_{i,I}$ and $K\gets K_{i,I}$ from \eqref{eq:algorithm-interval-kernel}.\\
\pc\>\>\>\textbf{for} $k=0,\ldots,m$ \textbf{do}\\
\pc\>\>\>\>Set $c_{i,k}\gets c_{i,k}+|I|(\Gamma-u h_f)$.\\
\pc\>\>\>\>\textbf{if} $k<m$ \textbf{then} $u\gets uK$.\\
\pc\>\>\>\textbf{end for}\\
\pc\>\>\textbf{end for}\\
\pc\>\>Define $d_i(x)=\Gamma w_i-A_i(x)$ using \eqref{eq:algorithm-bernstein}.\\
\pc\>\textbf{end for}\\
\pc\>Set $d_\bot(x)\gets0$ and $N\gets\lceil mj(j+1)/\epsilon\rceil$.\\
\pc\>Set $t_i\gets0$ for every $i\in L\cup\{\bot\}$.\\
\pc\>\textbf{repeat} $N$ times\\
\pc\>\>Choose $i$ maximizing $d_i(t_i/N)$, using the fixed tie order.\\
\pc\>\>Set $t_i\gets t_i+1$.\\
\pc\>\textbf{end repeat}\\
\pc\>\textbf{return} $v_i=t_i/N$ for every $i\in L\cup\{\bot\}$.
\end{pseudocode}

\begin{samepage}
Once $v$ is chosen, commit
$\nu_{i,I}\gets\nu_{i,I}R_{i,I}(v_i)^m$ for every stored interval.
This can be evaluated by $m$ successive sparse mixture updates.
The grid increments themselves are only hypothetical; they are not
committed as separate exposures. The committed update averages over
$\Bin(m,v_i)$ and does not use the candidate count that will actually
be sampled.
\end{samepage}

\begin{samepage}
\subsection{From candidates to one retained edge}
\label{sec:algorithm-assignment}

Draw $m$ independent candidates from $v$, using fresh randomness
independent of every seed. Let $r_i$ be the number of occurrences of
$i$. Each distinct candidate consumes $r_i$ consecutive exposures of
its own sampled seed. On a reset, the score is $W_f(Z_i)$ at the
age \emph{before} the reset, and the new age is zero. On a nonreset,
the score is $b_f$ and the age increments, capped at $C$. If these
scores are $X_{i,1},\ldots,X_{i,r_i}$, their combined desired win probability is
\[
                        Q_i=1-\prod_{k=1}^{r_i}(1-X_{i,k}).
\]
The scores have already been scaled individually. No auxiliary
success coins are generated: the algorithm only computes this product.
\end{samepage}

\begin{samepage}
If only one distinct candidate was sampled, it wins. Otherwise,
each candidate independently draws a bin from $K_{r_i}(Q_i,\cdot)$.
Only candidates in the highest drawn bin can win. The precomputed
within-bin probabilities determine which one does. Equal bin indices
need not give equal winner probabilities when the multiplicities
differ; Appendix~\ref{sec:explicit-tournament-preprocessing} gives
their exact formula. Algorithm~\ref{alg:finite-tournament} gives the
procedure $\mathrm{Tournament}$ used by Algorithm~\ref{alg:finite-matching}.
The numbers $Q_i$ are inputs to this calibrated procedure, rather
than probabilities to be normalized across candidates.
\end{samepage}

The dummy participates with its own persistent seed and the same
tournament rule; a dummy win discards the arrival. A real winner
receives the arrival, and free disposal keeps the heavier of its
new edge and its previously retained edge. All sampled exposures are
consumed, including those of losing candidates and those on rounds
whose winner is certain or whose edge cannot improve the retained
weight. They affect later selection opportunities.

\begin{pseudocode}{Finite online matching with free disposal}{alg:finite-matching}
\pc\>\textbf{Input:} offline vertices $L$ and the preprocessed finite model.\\
\pc\>Draw $Z_i\sim\pi_f$ independently for all $i\in L\cup\{\bot\}$.\\
\pc\>For each real $i$, set $\mathcal I_i\gets\varnothing$, $M_i\gets0$, and $e_i\gets\varnothing$.\\
\pc\>\textbf{for each} arrival $j=1,2,\ldots$ \textbf{do}\\
\pc\>\>Read weights $w_i$; set $w_i=0$ for an absent edge.\\
\pc\>\>For each real $i$, apply $\mathrm{Refine}(i,w_i)$.\\
\pc\>\>Compute $v$ by Algorithm~\ref{alg:probabilities} from these interval vectors.\\
\pc\>\>For every $i,I$, set $\nu_{i,I}\gets\nu_{i,I}R_{i,I}(v_i)^m$.\\
\pc\>\>Draw $m$ independent candidates from $v$; record their counts $r_i$.\\
\pc\>\>\textbf{for each} distinct candidate $i$ with $r_i>0$ \textbf{do}\\
\pc\>\>\>Set $F_i\gets1$.\\
\pc\>\>\>\textbf{repeat} $r_i$ times\\
\pc\>\>\>\>Draw a fresh reset bit $B\sim\mathrm{Bernoulli}(p)$.\\
\pc\>\>\>\>\textbf{if} $B=1$ \textbf{then}\\
\pc\>\>\>\>\>Set $X\gets W_f(Z_i)$, then $Z_i\gets0$.\\
\pc\>\>\>\>\textbf{else}\\
\pc\>\>\>\>\>Set $X\gets b_f$, then $Z_i\gets\min\{Z_i+1,C\}$.\\
\pc\>\>\>\>\textbf{end if}\\
\pc\>\>\>\>Set $F_i\gets F_i(1-X)$.\\
\pc\>\>\>\textbf{end repeat}\\
\pc\>\>\>Set $Q_i\gets1-F_i$.\\
\pc\>\>\textbf{end for}\\
\pc\>\>Set $u\gets\mathrm{Tournament}((i,r_i,Q_i)_{r_i>0})$.\\
\pc\>\>\textbf{if} $u\ne\bot$ and $w_u>M_u$ \textbf{then}\\
\pc\>\>\>Set $M_u\gets w_u$ and $e_u\gets(u,j)$, discarding its previous edge.\\
\pc\>\>\textbf{end if}\\
\pc\>\textbf{end for}\\
\pc\>\textbf{Output:} the nonempty retained edges $(e_i)_{i\in L}$.
\end{pseudocode}

\paragraph{Information used by the algorithm.}
The sampling routine reads only revealed weights and deterministic
accounting vectors. Neither the sampled ages nor the retained edges
are fed back into it. Likewise, its accounting vectors are not
conditioned on the realized multiplicities or the winner. For every
fixed input, the entire sequence of probability vectors is therefore
deterministic, exactly as required in Section~\ref{sec:selection}.
The auxiliary success trials and the independent geometric horizon
used to interpret $h_f$ belong only to the analysis.

\paragraph{Arithmetic and running time.}
All operations shown above can use exact rational arithmetic. With
$n=|L|$ and $d=C+1$, the credit coefficients and committed updates at
arrival $j$ require $O(njmd)$ arithmetic operations because the kernels
are sparse. Heap-based water filling requires
$O(N_j(m+\log(n+1)))$ further operations. The interval vectors and
credit coefficients use $O(njd+nm)$ space, in addition to the fixed
tournament tables. Appendix~\ref{sec:finite-implementation} gives the
polynomial bit-complexity bound and exact rational sampling procedure.
The algorithm has competitive guarantee $\Gamma-\epsilon$;
Section~\ref{sec:parameters} accounts for the cap, scaling, and grid
losses and obtains Theorem~\ref{thm:main}.

\section{A State-Dependent Primal--Dual Analysis}\label{sec:state-duality}

We now prove the matching reduction. Throughout this section, $W$
is any nondecreasing map from $\{-1,0,1,\ldots\}$ to $[0,1]$
satisfying Lemma~\ref{lem:tournament}. We write the formulas for
the uncapped process; Appendix~\ref{sec:finite-implementation}
gives the corresponding capped process and shows that the same
invariant and proof apply.

Fix an integer $m\ge2$ and $p\in(0,1)$, and put $q=1-p$
and $b=W(-1)$.
The age transition matrix and the substochastic failure matrix are
\[
 P(z,z+1)=q,\quad P(z,0)=p,
 \qquad
 T(z,z+1)=q(1-b),\quad T(z,0)=p(1-W(z)).
\]
Thus
\[
 T=(1-b)P-p(W-b)\delta_0.
\]
The stationary distribution of $P$ is $\pi(z)=pq^z$.  Each offline vertex has
an independent stationary seed; a reset outputs the age $z$ and a
nonreset outputs $-1$.  Its desired win probability is $W(z)$ at a
reset and $b$ at a nonreset.

The assumptions of Lemma~\ref{lem:tournament} give the
arbitrary-subset bound \eqref{eq:arbitrary-subset-proxy}. We use
the corresponding auxiliary miss process to lower-bound the
expected matching value.

Define the column vector
\begin{equation}\label{eq:age-h-resolvent}
 h=\mathbf1-(I-m(T-I))^{-1}\mathbf1,
 \qquad \gamma=\pi h.
\end{equation}
The inverse is well defined by
\[
 (I-m(T-I))^{-1}
 =\frac1{m+1}\sum_{n=0}^{\infty}
       \left(\frac m{m+1}\right)^nT^n.
\]
Equivalently, take an independent geometric number $N$ of
exposures, with $\Pr(N=n)=m^n/(m+1)^{n+1}$ for $n\ge0$.
Then $h(z)$ is the probability of an auxiliary success within
those $N$ exposures, starting at age $z$.
It follows that $0\le h\le1$ and that $h$ is nondecreasing in age.
For the latter assertion, couple two initial ages until the first
reset, using identical baseline-success coins at nonresets.  The
older process has a larger success probability at the first reset;
if both survive, their ages agree thereafter.

\begin{theorem}[State-dependent matching reduction]\label{thm:direct-age}
Under the assumptions above, there is a randomized online algorithm
for edge-weighted bipartite matching with free disposal whose
competitive ratio is at least
\[
 \gamma=1-\sum_{n=0}^{\infty}
       \frac{m^n}{(m+1)^{n+1}}\,\pi T^n\mathbf1.
\]
\end{theorem}

The standard matching LP gives the following sufficient condition.
If nonnegative variables satisfy $\alpha_i+\beta_j\ge\gamma w_{ij}$
for every edge and $\E[\ALG]\ge\sum_i\alpha_i+\sum_j\beta_j$,
then weak duality implies $\E[\ALG]\ge\gamma\OPT$.
We construct these variables online. Offline variables only increase;
the variable for an arrival is fixed when that arrival is processed.

\subsection{The state invariant}

At a fixed offline vertex and weight threshold $w$, let $\nu$ be the
subprobability row vector of the current age jointly with the event
that all auxiliary trials for edges of weight at least $w$ have failed. Initially
$\nu=\pi$.  Set $\theta=\pi-\nu$.

The invariant requires $\nu$ to be nonnegative and $\theta$ to be a
nonnegative linear combination of the distributions
\[
 \pi,\qquad \delta_0P^n\quad(n=0,1,2,\ldots).
\]
In particular, every state satisfying the invariant has $0\le\nu\le\pi$
coordinatewise. The coefficients need not sum to one;
their sum is the subprobability mass of $\theta$.  When $b=0$, a
probabilistic interpretation decomposes successful histories by their
first auxiliary success on an edge of weight at least $w$, which resets
the age to zero. For $b>0$, the
following algebraic invariant also includes a stationary component.

The following identities prove preservation of the invariant. For
an arrival whose edge to the fixed vertex has weight $w_i$, each
exposure with $w>w_i$ uses $P$ and maps
\[
 \nu\longmapsto\nu P,\qquad
 \theta\longmapsto\theta P.
\]
An exposure with $w\le w_i$ uses $T$ and maps
\[
 \nu\longmapsto\nu T,\qquad
 \theta\longmapsto b\pi+(1-b)\theta P+
                    p\left(\sum_z\nu(z)(W(z)-b)\right)\delta_0.
\]
Every coefficient is nonnegative because $W(z)\ge b$.
Both operations preserve the invariant, as do their convex combinations
with the identity operation.  This includes Bernoulli exposure trials
and arbitrary compositions of them.

$\delta_0P^n$ is the distribution of $\min(G,n)$ for a geometric
random variable $G$ of parameter $p$.  These distributions increase
stochastically with $n$, while $\pi P=\pi$.  Since $h$ is increasing
and nonnegative, $\theta Ph\ge\theta h$.  Define the credit
\[
 A(\nu):=(\pi-\nu)h
\]
and note that $A(\nu)\le\pi h=\gamma$. A $P$-update increases this credit because
$\theta Ph\ge\theta h$, or equivalently $\nu Ph\le\nu h$.
A $T$-update satisfies the same comparison: since
$0\le T\le P$ and $h\ge0$,
\[
                       \nu Th\le\nu Ph\le\nu h.
\]
Thus credit is nondecreasing under both types of exposure.
Consequently the residual
\[
 a(\nu):=\nu h=\gamma-A(\nu)
\]
is nonincreasing under either operation.  This is an assertion about
aggregate states satisfying the invariant; the pointwise inequality $Th\le h$ is neither
needed nor generally true.

\subsection{Updates for one online arrival}

Given a probability vector $v=(v_i)$, sample $m$ independent candidates and
pass their multiset to the tournament.  The multiplicity of a fixed
vertex $i$ is distributed as $\operatorname{Bin}(m,v_i)$.

For a current edge of weight $w_i$ and a threshold $w$, each exposure
uses $T$ if $w\le w_i$ and $P$ if $w>w_i$.
Writing $x=v_i$, the corresponding deterministic updates of the
joint state vector are
\[
 \nu\longmapsto \nu R_x^m,
 \quad R_x=I+x(T-I),\qquad w\le w_i,
\]
and
\[
 \nu\longmapsto \nu S_x^m,
 \quad S_x=I+x(P-I),\qquad w>w_i.
\]
These formulas already average over the binomial multiplicity.
Every intermediate state satisfies the invariant. The resulting credit is
continuous and nondecreasing as a function of $x\in[0,1]$.
One way to see monotonicity in $x$ is to condition on the number of
exposures: credit increases with that number, and the binomial count
increases stochastically with $x$.

Define the auxiliary failure probability and offline dual contribution by
\[
 y_i(w)=\nu_i(w)\mathbf1,
 \qquad
 \alpha_i(w)=\gamma-\nu_i(w)h,
 \qquad
 \alpha_i=\int_0^\infty\alpha_i(w)\,dw.
\]
The surrogate primal objective is
$\mathcal P=\sum_i\int_0^\infty(1-y_i(w))\,dw$.
The nonselection bound in \eqref{eq:arbitrary-subset-proxy} implies that the expected actual
matching value is at least $\mathcal P$.  Above all previously revealed
weights, $\nu_i(w)=\pi$, so the integrals have finite support.

Here is an explicit water-filling rule.  For each real vertex $i$,
let $\alpha_i[x]$ be its total offline credit after a hypothetical
update with probability $x$, and put
\[
 d_i(x)=\gamma w_i-\alpha_i[x],\qquad 0\le x\le1.
\]
Introduce a dummy vertex $\bot$ with $d_\bot(x)=0$.  Choose a maximizer, with a fixed deterministic tie rule, of the concave
optimization problem
\begin{equation}\label{eq:age-waterfill}
 \max\left\{\sum_{i\in L\cup\{\bot\}}
                 \int_0^{v_i}d_i(x)\,dx:
            v_i\ge0,\ \sum_i v_i=1\right\}.
\end{equation}
Each $d_i$ is continuous and nonincreasing, so a maximizer exists.
The first-order optimality conditions say that all positive
coordinates have a common terminal deficit $\beta$, while unused
coordinates have deficit at most $\beta$.  The dummy coordinate
ensures $\beta\ge0$.  Hence, after the updates,
\[
 \beta=\max\left(0,\max_i(\gamma w_i-\alpha_i)\right)
       =\sum_i v_i(\gamma w_i-\alpha_i).
\]
The maximization can equivalently be implemented by increasing the
coordinates with the largest current deficit until the total mass
is one, using the dummy coordinate whenever every real deficit is
nonpositive.  This establishes dual feasibility for the $\gamma$-scaled weights, which
persists because every $\alpha_i$ is nondecreasing.

\paragraph{Relation to the finite algorithm.}
The ideal algorithm uses this water-filling vector in the
sampling-and-tournament procedure of Section~\ref{sec:algorithm}.
The pseudocode there uses the capped, scaled seed and a rational
grid; Appendix~\ref{sec:finite-implementation} accounts for those
three implementation losses. We first prove the budget for the
ideal water-filling rule above.

\subsection{Thresholds at or below the current edge weight}

For $w\le w_i$, fix a pre-arrival state $\nu$ satisfying the invariant and write
\[
 y(x)=\nu R_x^m\mathbf1,\qquad a(x)=\nu R_x^m h.
\]
The resolvent identity in \eqref{eq:age-h-resolvent} is
\begin{equation}\label{eq:age-resolvent-identity}
 m(T-I)(h-\mathbf1)=h.
\end{equation}
Differentiating the finite polynomial $R_x^m$ gives
\[
 -y'(x)+a'(x)
 =m\nu R_x^{m-1}(T-I)(h-\mathbf1)
 =\nu R_x^{m-1}h
 \ge \nu R_x^mh=a(x).
\]
The inequality uses monotonicity on states satisfying the invariant for the final
Bernoulli exposure trial.  Integrating from zero to the chosen
probability $v_i$, and using that $a$ decreases, gives
\begin{equation}\label{eq:age-good-budget}
 y(0)-y(v_i)+a(v_i)-a(0)\ge v_i a(v_i).
\end{equation}
Thus the increase in the surrogate primal objective pays the offline-credit increase and the
appropriate share of the new online dual variable.

\subsection{Thresholds above the current edge weight}

For $w>w_i$, write $a(x)=\nu S_x^m h$. Define
\[
 k=(I-m(P-I))h.
\]
By \eqref{eq:age-resolvent-identity},
\begin{equation}\label{eq:age-nonnegative-k}
 k=m(P-T)(\mathbf1-h)
   =m\bigl[q b(1-h(z+1))+pW(z)(1-h(0))\bigr]
       \quad\text{at age }z.
\end{equation}
Hence $k\ge0$ pointwise.  Stationarity also gives $\pi k=\pi h=\gamma$.
Let $\rho=\nu S_x^{m-1}$.  Since $0\le\rho\le\pi$,
\[
 -a'(x)=-m\rho(P-I)h=\rho k-\rho h
 \le\gamma-\rho h
 \le\gamma-a(x).
\]
The final comparison again follows from monotonicity on states satisfying the invariant.
Integrating and using that $a$ decreases gives
\begin{equation}\label{eq:age-bad-budget}
 a(0)-a(v_i)\le v_i(\gamma-a(v_i)).
\end{equation}
There is no primal gain when $w>w_i$. The offline
dual-credit increase is bounded by $v_i$ times its post-arrival credit,
the right side of \eqref{eq:age-bad-budget}.  This exact inequality replaces the scalar gap-rate
condition.

\subsection{Completion of the competitive analysis}

For each vertex, integrate \eqref{eq:age-good-budget} below the current
edge weight and \eqref{eq:age-bad-budget} above it.  Summing over
vertices yields the following bound, where $\alpha_i(w)$ and
$\alpha_i$ denote their post-arrival values:
\begin{align*}
 \Delta\mathcal P-\sum_i\Delta\alpha_i
 &\ge\sum_i v_i\left[
       \int_0^{w_i}(\gamma-\alpha_i(w))\,dw
       -\int_{w_i}^{\infty}\alpha_i(w)\,dw\right]\\
 &=\sum_i v_i(\gamma w_i-\alpha_i)=\beta.
\end{align*}
All dual variables start at zero.  Therefore
$\mathcal P\ge\sum_i\alpha_i+\sum_j\beta_j$, while the dual constraints give
$\sum_i\alpha_i+\sum_j\beta_j\ge\gamma\,\mathrm{OPT}$.
Since the expected actual value is at least $\mathcal P$, the theorem follows.

\subsection{The choice of residual potential}
\label{sec:potential-optimality}

The resolvent also explains which part of the construction remains
to be optimized. For a fixed miss kernel, it gives the largest
stationary credit permitted by the local update inequality for $w\le w_i$ among
bounded linear residual potentials.

\begin{proposition}[Optimality for the local linear-credit criterion]
\label{prop:potential-optimality}
Let $g$ be a bounded column vector satisfying
\begin{equation}\label{eq:alternative-local-credit}
                  m\nu(T-I)(g-\one)\ge\nu g
                  \quad\text{for every state $\nu$ satisfying the invariant.}
\end{equation}
Then $\pi g\le\pi h=\gamma$. The vector $h$ attains equality in
\eqref{eq:alternative-local-credit} for every state.
\end{proposition}

\begin{proof}
Put $\mathcal R=(I-m(T-I))^{-1}$ and
$d=m(T-I)(g-\one)-g$. Every row $\pi T^n$ satisfies the invariant:
it is the state after $n$ successive $T$-updates. Thus $\pi T^nd\ge0$ for
every $n\ge0$. The positive resolvent series gives
\[
 0\le\frac1{m+1}\sum_{n\ge0}
       \left(\frac m{m+1}\right)^n\pi T^nd
   =\pi\mathcal R d
   =\pi(h-g).
\]
All series converge absolutely because the vectors are bounded.
The last identity and the equality assertion follow from
\eqref{eq:age-resolvent-identity}.
\end{proof}

The proposition concerns the local criterion
\eqref{eq:alternative-local-credit}. It places no upper bound on
matching algorithms or on other dual analyses. Within this criterion,
improving $\gamma$ requires changing the failure matrix.
We next optimize the win map that determines this kernel.

\section{Optimal Win Maps at a Fixed Reset Probability}
\label{sec:paper-pooling-theorem}

Fix $m\ge2$, $p\in(0,1)$, and $q=1-p$. The output $Y$ of the
stationary reset process has probabilities $\Pr(Y=-1)=q$ and
$\Pr(Y=z)=p^2q^z$ for $z\ge0$.  For a nondecreasing map
$W:\{-1,0,1,\ldots\}\to[0,1]$, let $T_W$ be its miss kernel and
define
\[
 \gamma(W)=1-(1-a)\sum_{n\ge0}a^n\pi T_W^n\mathbf1,
 \qquad a=\frac m{m+1},\qquad \pi(z)=pq^z.
\]
This section optimizes $\gamma(W)$ under the multiplicity-one
tournament constraint.  Feasibility for larger multiplicities is a
separate condition.
Let $\mathcal W$ be the class of nondecreasing maps
$W:\{-1,0,1,\ldots\}\to[0,1]$ satisfying
$W(Y)\preceq_{\mathrm{icx}}U^{m-1}$ for a uniform $U\in[0,1]$.

\begin{theorem}[Optimality of pooling]
\label{thm:paper-pooling}
Put $u_z=1-pq^z$ and $b_0=q^{m-1}/m$.  For
$b\in[b_0,1/m]$, define
\[
 \phi_b(u)=\max\{bu,u^m/m\},\qquad W^b(-1)=b,
 \qquad
 W^b(z)=\frac{\phi_b(u_{z+1})-\phi_b(u_z)}{u_{z+1}-u_z}.
\]
Every $W^b$ is nondecreasing, has stationary mean $1/m$, and satisfies
$W^b(Y)\preceq_{\mathrm{icx}}U^{m-1}$ for a uniform $U\in[0,1]$.
Moreover,
\[
\sup_{W\in\mathcal W}\gamma(W)
       =\max_{b_0\le b\le1/m}\gamma(W^b).
\]
Thus an optimal map exists and pools the youngest ages, possibly
modifying one boundary age, while preserving all older quantile-bin
scores. The statement concerns $\gamma(W)$ and this geometric
seed; it does not assert that pooling preserves the tournament
constraints for larger multiplicities.
\end{theorem}

We first establish the comparison principle that connects quantiles to
the online objective.

\begin{lemma}[Upper-tail comparison]
\label{lem:paper-upper-tail}
Let $W^0,W^1$ be nondecreasing maps with the same baseline
$W^0(-1)=W^1(-1)=b<1$.  If
\[
 \sum_{z\ge k}p^2q^zW^1(z)\ge
 \sum_{z\ge k}p^2q^zW^0(z)\qquad(k\ge0),
\]
then $\gamma(W^1)\ge\gamma(W^0)$.
\end{lemma}

\begin{proof}
We use a short renewal formula.  For a map $W$ with mean
$\mu=\E[W(Y)]$, write
\[
 r=aq(1-b),\quad t=r/q=a(1-b),\quad
 F_W(x)=\sum_{z\ge0}x^z(1-W(z)),\quad R=1-apF_W(r).
\]
Let $h=\mathbf1-(I-m(T_W-I))^{-1}\mathbf1$ and
$\ell_0=1-h(0)$.  Conditioning on the first reset, or summing the
miss recurrence from age zero, gives
\begin{equation}\label{eq:paper-pooling-renewal}
 \ell_0=\frac{1-a}{(1-r)R},\qquad
 \gamma(W)=\frac{m}{1+mb}\bigl[b+(\mu-b)\ell_0\bigr].
\end{equation}
For the second identity, stationarity implies
$\pi(P-T_W)=b\pi+(\mu-b)\delta_0$; apply this equality to
$h$ and use $m(I-T_W)(\mathbf1-h)=h$.
All denominators are positive because
\[
 R\ge\frac{1-r-ap}{1-r}>0,
 \qquad r+ap=a\bigl(q(1-b)+p\bigr)<1.
\]

Interpolate linearly between the two maps.  Regarding
$c_z=p^2q^zW(z)$ as the stationary score mass at age $z$,
differentiation of \eqref{eq:paper-pooling-renewal} gives
\begin{equation}\label{eq:paper-pooling-gradient}
 g_z:=\frac{\partial\gamma}{\partial c_z}
   =\frac{m\ell_0}{1+mb}
       \left(1-\frac{a(\mu-b)}{pR}\,t^z\right).
\end{equation}
Since $\mu\ge b$ and $0<t<1$, these marginal values are nondecreasing
in age.  They are positive as well.  Indeed,
$\mu=p+qb-p^2F_W(q)$, so the numerator of the smallest bracket is
\[
 R-\frac{a(\mu-b)}p
       =1-a(1-b)+ap\bigl(F_W(q)-F_W(r)\bigr)>0.
\]

Put $\delta_z=p^2q^z(W^1(z)-W^0(z))$ and
$D_k=\sum_{z\ge k}\delta_z\ge0$.  Along the interpolation,
summation by parts yields
\[
 \frac{d}{d\tau}\gamma((1-\tau)W^0+\tau W^1)
      =g_0D_0+\sum_{k\ge1}(g_k-g_{k-1})D_k\ge0.
\]
The series and differentiation are justified by the geometric bound on
the masses, the uniform positive lower bound on $R$, and the uniform
bound on the gradients.  Integrating in $\tau$ proves the lemma.
\end{proof}

\begin{proof}[Proof of Theorem~\ref{thm:paper-pooling}]
For a bounded random variable $X$ with quantile function $Q_X$, let
$J_X(u)=\int_u^1Q_X(v)\,dv$.  The elementary dual identities
\[
 \begin{split}
 J_X(u)&=\inf_s\{(1-u)s+\E[(X-s)_+]\},\\
 \E[(X-s)_+]&=\sup_{0\le u\le1}\{J_X(u)-(1-u)s\}
 \end{split}
\]
show that increasing-convex order is equivalent to comparison of these
integrated upper quantiles.  In particular, the reference variable
$V=U^{m-1}$ has mean $1/m$ and
$J_V(u)=(1-u^m)/m$.

The convex function $\phi_b$ has $\phi_b(0)=0$,
$\phi_b(q)=bq$, and $\phi_b(1)=1/m$, with all slopes in $[0,1]$.
Its successive secants therefore define a nondecreasing $W^b$ in
$[0,1]$.  The integrated lower-quantile curve of $W^b(Y)$ is the
chordal interpolant of $\phi_b$ at
$0,q,u_1,u_2,\ldots,1$.  Each chord lies above $\phi_b$, which lies
above $u^m/m$.  The endpoint at one gives the full mean $1/m$, so its
upper-quantile curve lies below $J_V$.  This proves singleton
feasibility.  The crossing of the line $bu$ with $u^m/m$ also gives
the stated pooling form; at most one age bin contains the crossing.

Now let $W$ be any map in the supremum, and set
$b=W(-1)$ and $\mu=\E[W(Y)]$.  Monotonicity and singleton
feasibility imply $b\le\mu\le1/m$.  At the atom cut $u_k$, the
integrated upper tail is
\[
 U_k=\sum_{z\ge k}p^2q^zW(z).
\]
The quantile constraint gives $U_k\le(1-u_k^m)/m$.
All lower atoms have score at least $b$ and total mass $u_k$, so
$U_k\le\mu-bu_k\le1/m-bu_k$ as well.
If $b\ge b_0$, these bounds combine to give
\[
 U_k\le\frac1m-\max\{bu_k,u_k^m/m\}
       =\sum_{z\ge k}p^2q^zW^b(z).
\]
Lemma~\ref{lem:paper-upper-tail} proves
$\gamma(W)\le\gamma(W^b)$, including competitors with mean below
$1/m$.

If $b<b_0$, keep this baseline and replace the reset scores by the
unpooled quantile-bin means.  Their upper tails are exactly
$(1-u_k^m)/m$, so the same lemma shows that this replacement cannot
lower $\gamma$.  Raising its baseline to $b_0$ then increases every
desired win probability on every seed path and cannot lower the discounted
success probability.  The resulting map is $W^{b_0}$.

Finally, the maximum on the right is attained.  For each fixed age,
$W^b(z)$ is continuous in $b$, including at pooling boundaries and
at $b=1/m$.  Bounded convergence makes
$\pi T_{W^b}^n\mathbf1$ continuous for every fixed $n$, and the
resolvent tail is bounded uniformly in $b$ by $a^{N+1}$ after index
$N$.  Hence $\gamma(W^b)$ is continuous on the compact interval
$[b_0,1/m]$.  Every map in this interval is singleton-feasible, which
also proves the reverse inequality in the theorem.
\end{proof}

\begin{remark}
The competitors include every bounded monotone map on the infinite age
space, including capped and other finite-valued maps.  If one restricts
the optimization to finite-valued maps, the same value is a supremum:
pooling a sufficiently far upper tail of $W^b$ into its conditional
mean preserves singleton feasibility, and the resulting values of
$\gamma$ converge to $\gamma(W^b)$.
Attainment by a finite-valued map is not required or asserted.
\end{remark}

\subsection{Why the age state suffices}
\label{sec:regenerative-purification}

The preceding optimization also covers seeds with a richer internal
state. The relevant property is that a reset cycle produces at most one
output above the fixed baseline. Averaging that output over the
internal randomness leaves the avoidance objective unchanged.

Let $\mathcal S$ be a finite or countable partially ordered state
space with a least element $0$. A Markov kernel $R$ on $\mathcal S$
is \emph{stochastically monotone} if $Rf$ is increasing whenever
$f$ is bounded and increasing. Write $\mu\preceq_{\st}\nu$ if
$\mu f\le\nu f$ for every such $f$. Fix such a kernel, a reset probability
$p\in(0,1)$, a baseline $b\in[0,1)$, and an increasing function
$W:\mathcal S\to[b,1]$.

On each exposure, independently of the past, the seed resets with
probability $p$. A reset from state $Z$ outputs $W(Z)$ and sends the
state to $0$. Otherwise, the output is $b$ and the next state is
sampled from $R(Z,\cdot)$, using fresh randomness. Put $q=1-p$.
The transition kernel and stationary distribution are
\[
 \begin{gathered}
 P=p\one\delta_0+qR,
 \qquad T=p(\one-W)\delta_0+q(1-b)R,\\
 \pi=p\sum_{k\ge0}q^k\delta_0R^k.
 \end{gathered}
\]
The stationary process can be constructed from independent reset bits
indexed by all integer times: start from $0$ at the last reset and
apply the intervening $R$ transitions.

\begin{proposition}[Reduction to age-dependent scores]
\label{prop:regenerative-purification}
For the stationary seed above, define
\[
 \overline W(-1)=b,\qquad
 \overline W(k)=\delta_0R^kW\quad(k\ge0).
\]
This is a nondecreasing geometric-age map with reset probability $p$.
Write $X_t$ and $\overline X_t$ for the original and age-dependent outputs.
For every finite deterministic set of exposure indices $S$,
\begin{equation}\label{eq:purification-products}
 \E\left[\prod_{t\in S}(1-X_t)\right]
       =\E\left[\prod_{t\in S}(1-\overline X_t)\right].
\end{equation}
Moreover, the combined desired win probabilities
$U_S=1-\prod_{t\in S}(1-X_t)$ and
$\overline U_S=1-\prod_{t\in S}(1-\overline X_t)$ admit a coupling
such that
\begin{equation}\label{eq:purification-conditional-mean}
                 \overline U_S=\E[U_S\mid\mathcal B],
\end{equation}
where $\mathcal B$ is the sigma-field generated by the two-sided
reset sequence. Thus averaging over the internal randomness preserves
every discounted sum of these failure probabilities, and all the tournament inequalities
of Lemma~\ref{lem:tournament}.
\end{proposition}

\begin{proof}
Condition on the complete reset sequence. Each cycle contains
independent internal $R$ randomness and has just one possibly
nonbaseline output, at its terminating reset. If that reset follows
$k$ ordinary transitions, the conditional mean of its output is
$\delta_0R^kW$. All other outputs equal $b$. Consequently,
\[
 \E\left[\prod_{t\in S}(1-X_t)\,\middle|\,\mathcal B\right]
   =\prod_{t\in S}\bigl(1-\E[X_t\mid\mathcal B]\bigr).
\]
Using the same reset bits in the geometric-age process gives
$\overline X_t=\E[X_t\mid\mathcal B]$ and proves both displayed
identities. This reasoning also applies to the cycle crossing the
start of $S$: its preceding reset is fixed by the two-sided sequence.

Since $0$ is the least state, $\delta_0\preceq_{\st}\delta_0R$.
Stochastic monotonicity then gives
$\delta_0R^k\preceq_{\st}\delta_0R^{k+1}$ for every $k$.
Applying $W$ proves that $\overline W$ is nondecreasing; its range
lies in $[b,1]$.

Equation~\eqref{eq:purification-products} preserves every term of
the absolutely convergent discounted avoidance sum. Conditional
Jensen applied to \eqref{eq:purification-conditional-mean} gives
$\E[\varphi(\overline U_S)]\le\E[\varphi(U_S)]$ for every convex
$\varphi:[0,1]\to\R$. Taking $S$ to be a consecutive block and
$\varphi(u)=(u-t)_+$ proves all the tournament comparisons.
\end{proof}

\begin{corollary}\label{cor:regenerative-pooled-optimality}
Fix $m$ and $p$. Among the monotone regenerative seeds above
satisfying the singleton constraint
$X_0\preceq_{\icx}U^{m-1}$, the maximum value of $\gamma$ in
\eqref{eq:intro-score} is
\[
                \max_{q^{m-1}/m\le b\le1/m}\gamma(W^b).
\]
In particular, a pooled geometric-age map attains the optimum.
\end{corollary}

\begin{proof}
The reduction to age-dependent scores preserves $\gamma$ and the singleton constraint,
so Theorem~\ref{thm:paper-pooling} gives the upper bound. The
geometric-age process itself belongs to this class: take
$\mathcal S=\mathbb Z_{\ge0}$ and $R(k,k+1)=1$. The maximum
is therefore attained by the pooled map from that theorem.
\end{proof}

The reduction requires no negative association of the original output;
the resulting nondecreasing age map has that property. The corollary
optimizes under the singleton constraint. Averaging preserves any
constraints for repeated candidates that already hold, whereas replacing
the resulting map by its pooled optimizer still requires checking those
additional constraints. Scores above the baseline on intermediate
transitions are outside the regenerative class considered here.

\section{Setting the Parameters}
\label{sec:parameters}

\begin{figure}[!htbp]
\centering
\begin{tikzpicture}
\begin{axis}[
 width=.84\linewidth,height=5.7cm,
 xmin=3.7,xmax=4.8,ymin=.0075,ymax=.026,
 xlabel={$x=500(1-u)$},ylabel={$500\,L(u)$},
 axis lines=left,
 xtick={3.7,4.0,4.3,4.6,4.8},
 ytick={.01,.015,.02,.025},
 scaled y ticks=false,
 yticklabel style={/pgf/number format/fixed,/pgf/number format/precision=3},
 tick label style={font=\small},
 label style={font=\small},
 legend style={at={(.98,.98)},anchor=north east,
   draw=none,fill=white,font=\small,cells={anchor=west}},
 clip=true]
\addplot[gray,dashed,line width=1pt] coordinates {
(3.7000000000,0.0243857115)
(3.7137500000,0.0240502325)
(3.7275000000,0.0237193597)
(3.7412500000,0.0233930299)
(3.7550000000,0.0230711809)
(3.7687500000,0.0227537513)
(3.7825000000,0.0224406805)
(3.7962500000,0.0221319087)
(3.8100000000,0.0218273771)
(3.8237500000,0.0215270276)
(3.8375000000,0.0212308027)
(3.8512500000,0.0209386461)
(3.8650000000,0.0206505019)
(3.8787500000,0.0203663151)
(3.8925000000,0.0200860315)
(3.9062500000,0.0198095976)
(3.9200000000,0.0195369606)
(3.9337500000,0.0192680685)
(3.9475000000,0.0190028699)
(3.9612500000,0.0187413142)
(3.9750000000,0.0184833515)
(3.9887500000,0.0182289325)
(4.0025000000,0.0179780085)
(4.0162500000,0.0177305318)
(4.0300000000,0.0174864550)
(4.0437500000,0.0172457315)
(4.0575000000,0.0170083153)
(4.0712500000,0.0167741611)
(4.0850000000,0.0165432242)
(4.0987500000,0.0163154604)
(4.1125000000,0.0160908262)
(4.1262500000,0.0158692787)
(4.1400000000,0.0156507756)
(4.1537500000,0.0154352751)
(4.1675000000,0.0152227361)
(4.1812500000,0.0150131179)
(4.1950000000,0.0148063804)
(4.2087500000,0.0146024842)
(4.2225000000,0.0144013903)
(4.2362500000,0.0142030603)
(4.2500000000,0.0140074561)
(4.2637500000,0.0138145405)
(4.2775000000,0.0136242766)
(4.2912500000,0.0134366279)
(4.3050000000,0.0132515587)
(4.3187500000,0.0130690335)
(4.3325000000,0.0128890174)
(4.3462500000,0.0127114760)
(4.3600000000,0.0125363753)
(4.3737500000,0.0123636819)
(4.3875000000,0.0121933627)
(4.4012500000,0.0120253852)
(4.4150000000,0.0118597172)
(4.4287500000,0.0116963270)
(4.4425000000,0.0115351833)
(4.4562500000,0.0113762555)
(4.4700000000,0.0112195129)
(4.4837500000,0.0110649257)
(4.4975000000,0.0109124643)
(4.5112500000,0.0107620995)
(4.5250000000,0.0106138025)
(4.5387500000,0.0104675449)
(4.5525000000,0.0103232987)
(4.5662500000,0.0101810365)
(4.5800000000,0.0100407308)
(4.5937500000,0.0099023548)
(4.6075000000,0.0097658821)
(4.6212500000,0.0096312866)
(4.6350000000,0.0094985424)
(4.6487500000,0.0093676242)
(4.6625000000,0.0092385069)
(4.6762500000,0.0091111657)
(4.6900000000,0.0089855763)
(4.7037500000,0.0088617146)
(4.7175000000,0.0087395569)
(4.7312500000,0.0086190798)
(4.7450000000,0.0085002603)
(4.7587500000,0.0083830755)
(4.7725000000,0.0082675031)
(4.7862500000,0.0081535208)
(4.8000000000,0.0080411069)
};
\addlegendentry{Reference $L(u)=u^{500}/500$}
\addplot[blue!70!black,line width=1.2pt] coordinates {
(3.5896646306,0.0272524370)
(3.6222650157,0.0263720765)
(3.6551614689,0.0255124899)
(3.6883566791,0.0246734397)
(3.7218533593,0.0238546842)
(3.7556542475,0.0230559775)
(3.7897621065,0.0222770698)
(3.8241797240,0.0215177076)
(3.8589099132,0.0207776339)
(3.8939555128,0.0200565882)
(3.9293193873,0.0193543069)
(3.9650044272,0.0186705232)
(4.0010135491,0.0180049678)
(4.0373496964,0.0173573684)
(4.0740158389,0.0167274507)
(4.1110149737,0.0161149377)
(4.8000000000,0.0160925477)
};
\addlegendentry{Pooled lower-quantile curve}
\addplot[only marks,mark=*,mark size=1.35pt,blue!70!black,
 forget plot] coordinates {
(4.3011123424,0.0161087601)
(4.2624023313,0.0161100180)
(4.2240407103,0.0161112647)
(4.1860243439,0.0161125001)
(4.1483501248,0.0161137244)
(4.1110149737,0.0161149377)
(4.0740158389,0.0167274507)
(4.0373496964,0.0173573684)
(4.0010135491,0.0180049678)
(3.9650044272,0.0186705232)
(3.9293193873,0.0193543069)
(3.8939555128,0.0200565882)
(3.8589099132,0.0207776339)
};
\draw[gray,densely dotted] (axis cs:4.1110149737,.0075)
 -- (axis cs:4.1110149737,.026);
\node[font=\small,rotate=90,anchor=south,text=gray!80!black]
 at (axis cs:4.1110149737,.0108) {$u=u_{10}$};
\node[font=\small,text=blue!70!black,align=center]
 at (axis cs:4.52,.0182) {Pooled bins};
\draw[->,blue!70!black] (axis cs:4.52,.0175)
 -- (axis cs:4.52,0.0161016469);
\node[font=\small,text=blue!70!black,align=center]
 at (axis cs:3.86,.0125) {Original tail bins};
\draw[->,blue!70!black] (axis cs:3.86,.0133)
 -- (axis cs:3.95,0.0189550449);
\end{axis}
\end{tikzpicture}
\caption{The pooled map for the simpler instantiation, with $m=500$,
$p=9/1000$, and $A=10$.  The horizontal coordinate reverses and
magnifies the upper quantiles; the vertical coordinate scales their
integral by $500$.  The blue curve is linear for $u\le u_{10}$
(right of the dotted line) and joins the unchanged quantile-bin endpoints
on the left. Dots show selected bin endpoints; the tail chords lie
slightly above the gray reference curve.}
\label{fig:quantile-pooling}
\end{figure}
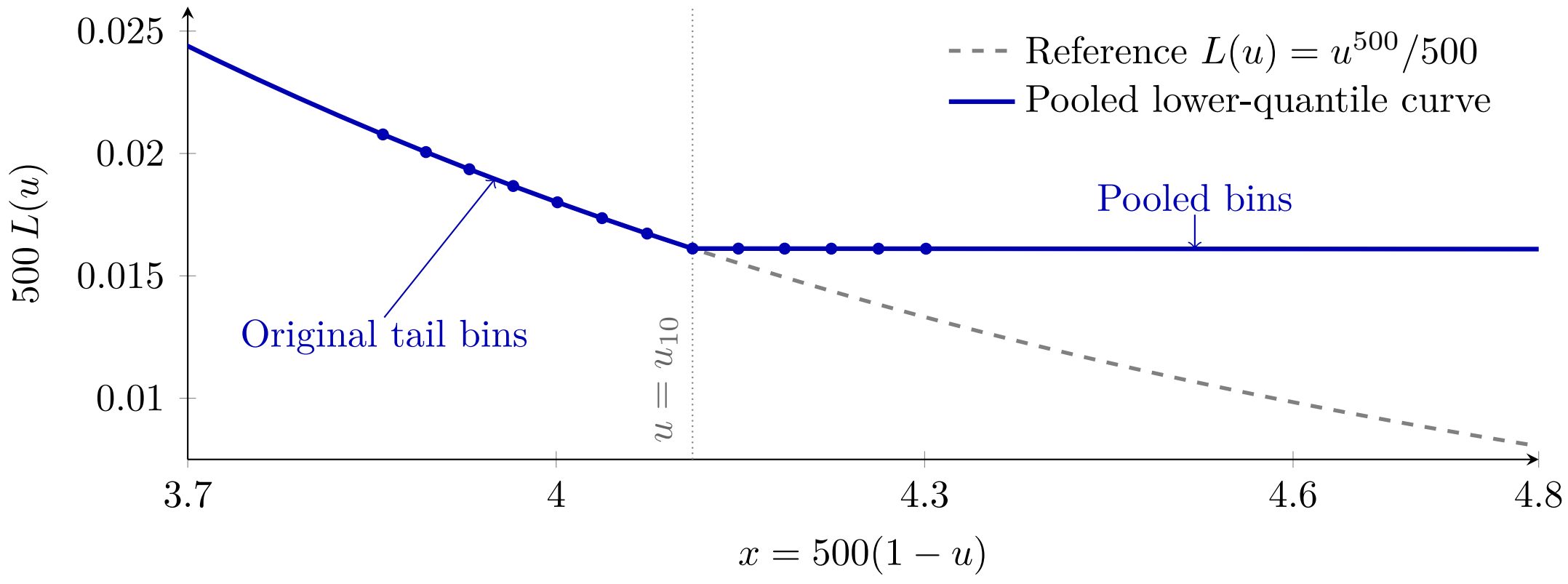

We use two instantiations of the following family:
\begin{equation}\label{eq:paper-parameters}
 m\in\{500,1000\},\qquad p=\frac9{2m},\qquad q=1-p,
 \qquad A=\frac m{50},\qquad
 b=\frac{(1-pq^A)^{m-1}}m.
\end{equation}
The choice $m=1000$ proves Theorem~\ref{thm:main}; the choice
$m=500$ gives a $0.556$-competitive construction with a shorter
certificate.
Put $u_z=1-pq^z$. The win map is
\begin{equation}\label{eq:paper-explicit-map}
 W(-1)=b,\qquad
 W(z)=
 \begin{cases}
 b,&0\le z<A,\\[2pt]
 \displaystyle\frac{u_{z+1}^m-u_z^m}{mp^2q^z},&z\ge A.
 \end{cases}
\end{equation}
It is the pooled map $W^b$ of Theorem~\ref{thm:paper-pooling} at an
integer pooling boundary, illustrated in Figure~\ref{fig:quantile-pooling}.
Thus it is nondecreasing, has mean $1/m$,
and satisfies the singleton tournament constraint.

\begin{table}[!htbp]
\centering
\small
\begin{tabular}{@{}rccccc@{}}
\toprule
$m$ & $p$ & $A$ & $s_0$ & Ideal lower bound $g_m$
    & Finite lower bound\\
\midrule
$500$ & $9/1000$ & $10$ & $10^{-6}$
      & $0.55607$ & $0.556068988$\\
$1000$ & $9/2000$ & $20$ & $10^{-7}$
      & $0.55619680$ & $0.556195789$\\
\bottomrule
\end{tabular}
\caption{Two certified instantiations. Here $s_0$ separates the
small-threshold estimate from the chord verification. Every displayed
lower bound is strict. The ideal bounds use the same $20$-term rational
Bonferroni expression; the finite bounds subtract
$10^{-8}+1/(m\cdot10^6)+10^{-6}$ for capping, scaling, and
water filling, respectively.}
\label{tab:paper-instantiations}
\end{table}

The proof uses the conditions summarized below. The selection
conditions concern the constant-size
random process and are independent of the input graph.

\begin{center}
\small
\begin{tabular}{@{}p{0.35\linewidth}p{0.58\linewidth}@{}}
\toprule
Condition & Verification \\
\midrule
Monotone scores in $[0,1]$
 & Quantile construction, Theorem~\ref{thm:paper-pooling}.\\[3pt]
Stationarity and negative association
 & Geometric-age process; Lemma~\ref{lem:paper-age-na}.\\[3pt]
Tournament condition for $r=1$
 & Conditional Jensen applied to $U^{m-1}$.\\[3pt]
Tournament condition for $2\le r< m$, all $s\in[0,1]$
 & Jensen and chord certificate of Lemma~\ref{lem:paper-numeric-tournament};
   all $498$ or $998$ required multiplicities are covered.\\[3pt]
Competitive guarantee $\gamma>g_m$
 & A $20$-term rational lower bound,
   Lemma~\ref{lem:paper-score} and Table~\ref{tab:paper-instantiations}.\\[3pt]
Implementation loss \newline $<1.012\cdot10^{-6}$
 & Age cap, rational tournament, and water filling,
   Appendix~\ref{sec:finite-implementation}.\\
\bottomrule
\end{tabular}
\end{center}

\subsection{The tournament inequalities}

\begin{lemma}[Computer-assisted tournament certificate]
\label{lem:paper-numeric-tournament}
For the parameters in \eqref{eq:paper-parameters} and the map
\eqref{eq:paper-explicit-map},
\[
 \E\left[\left(s-\prod_{k=1}^r(1-W(Y_k))\right)_+\right]
       \le H_r(s)
 \qquad(1\le r<m,\ 0\le s\le1).
\]
\end{lemma}

\begin{proof}
The singleton case follows analytically from the quantile construction.
Appendix~\ref{sec:certificate} reduces each remaining case to a finite
collection of explicit inequalities. Condition on the first reset;
two positive renewal recurrences give the minimum and mean failure
of the fresh remainder. Replacing that remainder by its endpoint
chord and integrating out the initial age by conditional Jensen
gives a convex upper bound with at most $2r$ terms and a no-reset
atom. We take the values of $s_0$ in
Table~\ref{tab:paper-instantiations} and compare this bound with the
reference by a quadratic estimate on $[0,s_0]$ and rational chords
on $[s_0,1]$. Support and mean bounds use $128$-bit dyadic rationals.
Failure scales are rounded downward to $64$ bits, mixture weights
upward to $128$ bits, and chord endpoint values upward to $96$ bits.
All interval evaluations use $256$ bits of precision.

The repository verifier described in Appendix~\ref{sec:verifier} passes for every
required multiplicity in both instantiations. Its certified statistics
are summarized below.
\begin{center}
\small
\begin{tabular}{@{}rcccc@{}}
\toprule
$m$ & Multiplicities & Chords & Small-threshold ratio
    & Global chord gap\\
\midrule
$500$ & $2,\ldots,499$ & $12{,}915$ & $<0.998527$
      & $>1.49\cdot10^{-19}$\\
$1000$ & $2,\ldots,999$ & $30{,}105$ & $<0.999116$
      & $>2.94\cdot10^{-22}$\\
\bottomrule
\end{tabular}
\end{center}
The ratio is the largest upper-to-lower coefficient ratio in the
small-threshold comparison. Each chord is certified on the entire
unit interval by a strict interval enclosure of the equivalent
integer-power comparison in \eqref{eq:paper-prefix-powers}.
Neither run required an exact-power fallback. The analytic bounds
and the checked partitions cover every real threshold in $[0,1]$.
\end{proof}

\subsection{A short rational lower bound on the guarantee}

$\gamma$ can be evaluated without truncating the
age space. We specialize the renewal formula to mean $1/m$.

\begin{lemma}[Renewal formula for $\gamma$]\label{lem:paper-flux}
Let $W$ be a geometric-age win map with baseline $b$ and mean $1/m$.
Set $a=m/(m+1)$, $\rho=aq(1-b)$, and
$F(v)=\sum_{z\ge0}v^z(1-W(z))$. Then
\begin{equation}\label{eq:paper-flux}
 \gamma=\frac{mb+(1-mb)\ell_0}{1+mb},\qquad
 \ell_0=\frac{1-a}{(1-\rho)(1-apF(\rho))}.
\end{equation}
\end{lemma}

\begin{proof}
This is \eqref{eq:paper-pooling-renewal} with $\mu=1/m$ and $r=\rho$.
\end{proof}

For an integer-pooled map, put $B=1-b$ and
\[
 c_j=(-1)^{j+1}\binom{m-1}{j}
                   \frac{p^{j-1}(1-q^{j+1})}{j+1}.
\]
The identity
\[
 1-W(z)=\frac1p\int_q^1
       \bigl[1-(1-pq^zv)^{m-1}\bigr]\,dv\qquad(z\ge A)
\]
shows that the even Bonferroni truncation
$\sum_{j=1}^{2k}c_jq^{jz}$ is a lower bound on $1-W(z)$.
In particular,
\begin{equation}\label{eq:paper-S20}
 S_{20}:=B\frac{1-\rho^A}{1-\rho}
       +\rho^A\sum_{j=1}^{20}
               \frac{c_jq^{Aj}}{1-\rho q^j}
       \le F(\rho).
\end{equation}
These bounds do not require successive expansion terms to decrease
in magnitude. They follow from the usual Bonferroni inequalities for
$1-(1-x)^{m-1}$, followed by integration.

\begin{lemma}\label{lem:paper-score}
The parameters in \eqref{eq:paper-parameters} satisfy
\[
 0<mb<1,\qquad 1-apS_{20}>0,
\]
and
\begin{equation}\label{eq:paper-rational-score}
 \frac{mb+(1-mb)\dfrac{1-a}
          {(1-\rho)(1-apS_{20})}}{1+mb}
       >g_m.
\end{equation}
Here $g_{500}=55607/100000$ and $g_{1000}=347623/625000$.
Consequently $\gamma>g_m$ for both rows of
Table~\ref{tab:paper-instantiations}.
\end{lemma}

\begin{proof}
All quantities in the displayed inequalities are rational expressions
in \eqref{eq:paper-parameters}. Exact rational evaluation verifies
them. The score verifier identified in Appendix~\ref{sec:score-verifier} evaluates
the displayed expression using rational arithmetic and compares it
with the corresponding exact rational $g_m$. Finally, the right side of
\eqref{eq:paper-flux} is increasing in $F(\rho)$ when $mb<1$.
Substituting \eqref{eq:paper-S20} proves the score bound.
\end{proof}

\subsection{Completing the competitive guarantee}

Theorem~\ref{thm:direct-age} and
Lemmas~\ref{lem:paper-numeric-tournament}--\ref{lem:paper-score}
give an ideal algorithm with competitive guarantee greater than $g_m$ for
each row of Table~\ref{tab:paper-instantiations}.
For a finite implementation, use the common choices
\[
 C=5m/2,\qquad
 \eta=1-\frac1{m\cdot10^6},\qquad
 \epsilon=10^{-6}.
\]
Use the seed map $W_f(z)=\eta W(\min\{z,C\})$
for $z\ge0$, with baseline $W_f(-1)=\eta b$, and
water-filling accuracy $\epsilon$.
Appendix~\ref{sec:finite-implementation} proves that these operations
lose less than $10^{-8}$ from capping, at most $1/(m\cdot10^6)$
from scaling, and at most $10^{-6}$ from water filling, while
preserving tournament feasibility. The resulting guarantees
are therefore strictly greater than
\[
 \begin{aligned}
 m=500:\quad&
 0.55607-10^{-8}-2\cdot10^{-9}-10^{-6}
       =0.556068988>0.556,\\
 m=1000:\quad&
 0.55619680-10^{-8}-10^{-9}-10^{-6}
       =0.556195789>0.556195.
 \end{aligned}
\]
The second instantiation proves Theorem~\ref{thm:main}.

\paragraph{Scope of the Lean development.}
The accompanying Lean development formalizes the $\BestRatio$ guarantee
for an explicit finite dyadic seed, relative to two explicitly registered
literature assumptions: a calibrated seed-to-selector construction and
an independent categorical sampler. These assumptions are not proved
in the development. The formal theorem does not cover the expected
polynomial running time, the ideal-algorithm tightness theorem, or the
optimization characterizations proved here. The formal statements,
assumption registry, and verification instructions are included with
the Lean sources in the artifact repository
(Appendix~\ref{sec:verifier}).

\noindent\textbf{AI Use Statement.} This work resulted from extensive
interaction between the authors and OpenAI Astra. The authors provided
suggestions, critically checked the responses, and directed the writing
and revision process, evaluating the proposed ideas and arguments.

\bibliographystyle{alpha}
\bibliography{online_weighted_matching}

\appendix

\begin{samepage}
\section{Negative Association of the Geometric-Age Process}
\label{sec:paper-geometric-na}

We give an independent proof of the seed property used in
Section~\ref{sec:selection}. It uses
Reimer's inequality \cite{Reimer00}: for arbitrary events $E,F$ in a
finite Bernoulli product space, with arbitrary coordinate biases,
\[
                 \Pr(E\mathbin\Box F)\le\Pr(E)\Pr(F).
\]
Here $E\mathbin\Box F$ means that disjoint sets of product coordinates
certify the two events.  No monotonicity in the product coordinates is
required.

\begin{lemma}\label{lem:paper-age-na}
The outputs of the stationary geometric-age process are negatively associated.
The same holds after any finite age cap.
\end{lemma}

\begin{proof}
Realize the seed using iid Bernoulli reset bits $(B_t)_{t\in\mathbb Z}$.
Fix a cap $L$.  If $B_t=0$, let $Y_t^{[L]}=-1$; otherwise let
$Y_t^{[L]}$ be the smaller of $L$ and the number of consecutive zeros
immediately preceding $t$.  Any finite set of capped outputs depends on
finitely many bits.

At a realized nonnegative output $Y_t^{[L]}=z$, the bits in
$K_t=[t-z,t]\cap\mathbb Z$ consist of $z$ zeros followed by one.
Fixing them certifies $Y_t^{[L]}\ge z$ under every completion.
An output equal to $-1$ needs no certificate.  All the sets $K_t$ of
nonnegative outputs in a realized sample are disjoint: if $s<t$ and their
intervals intersected, the terminal one at $s$ would lie among the
zeros required by the later interval.

Let $E,F$ be increasing events in disjoint sets of output coordinates.
At any sample in $E\cap F$, the certificates of its nonnegative outputs
in the first set jointly certify $E$, since every completion preserves
or increases those outputs.  The analogous certificates certify $F$,
and the two collections are disjoint.  Thus
$E\cap F=E\mathbin\Box F$ in the Bernoulli product space.
Reimer's inequality proves negative correlation of these events.
Integrating upper-level events proves the covariance inequality for
bounded increasing functions, which is negative association.

For any finite set of times, capped outputs eventually equal the
uncapped outputs almost surely as $L\to\infty$.  Bounded convergence
passes negative association to the uncapped sequence; truncation
extends it to the usual integrable functions.
\end{proof}
\end{samepage}

\section{A Finite Certificate for the Tournament Inequalities}
\label{sec:certificate}

We prove the reduction used in
Lemma~\ref{lem:paper-numeric-tournament}.
Fix the integer-pooled map \eqref{eq:paper-explicit-map}, and write
$D(z)=1-W(z)$ and $B=1-b$. For each $2\le r<m$, we prove
\begin{equation}\label{eq:certificate-stop-loss}
 \E[(s-V_r)_+]\le H_r(s),\qquad
 V_r=\prod_{j=1}^r(1-X_j),\qquad 0\le s\le1.
\end{equation}
The proof conditions on the first reset. Its bounded remainder is
replaced by a two-point distribution with the same mean; the unbounded
initial age is integrated out. The resulting convex envelope has
at most $2r$ terms for the initial age and a no-reset atom.

\subsection{The support and mean of a fresh block}

Let $V_n^0$ be the failure product in $n$ exposures started at age
zero, and put $d_n=\min V_n^0$ and $f_n=\E[V_n^0]$.
Then $d_0=f_0=1$, and conditioning on the first reset gives
\begin{align}
 d_n&=\min\left\{B^n,\min_{0\le k<n}
                         B^kD(k)d_{n-k-1}\right\},\nonumber\\
 f_n&=(qB)^n+p\sum_{k=0}^{n-1}(qB)^kD(k)f_{n-k-1}.
 \label{eq:paper-fresh-renewal}
\end{align}
Indeed, the no-reset event has probability $q^n$ and product $B^n$.
A first reset after $k$ nonresets has probability $pq^k$ and leaves
an independent fresh block of length $n-k-1$.
Every failure factor is at most $B$, so the maximum product is
$B^n$. In particular $d_n\le V_n^0\le B^n$.
For $n\le A$, every score is the baseline and $d_n=f_n=B^n$.

For $d_n<B^n$, set $\theta_n=(B^n-f_n)/(B^n-d_n)$; otherwise
set $\theta_n=0$. The endpoint chord of a convex function gives
\begin{equation}\label{eq:paper-fresh-chord}
 \E[(s-V_n^0)_+]
 \le(1-\theta_n)(s-B^n)_++\theta_n(s-d_n)_+.
\end{equation}
The two-point distribution has the same mean as $V_n^0$. These recurrences
and this comparison account for every reset pattern.

\subsection{Integrating out the initial age}

Put $u_k=1-pq^k$. For $C>0$ and $0\le u<1$, define
$I(s,C,u)=\int_u^1(s-C(1-v^{m-1}))_+\,dv$. Direct integration gives
\begin{equation}\label{eq:paper-jensen-integral}
 I(s,C,u)=
 \begin{cases}
 C H_1(s/C),&0\le s\le C(1-u^{m-1}),\\[2pt]
 (1-u)(s-C)+C(1-u^m)/m,&s\ge C(1-u^{m-1}).
 \end{cases}
\end{equation}
The branches agree in value and derivative. For the stationary
age $Z$, conditional Jensen gives
\begin{equation}\label{eq:paper-jensen-pattern}
 \E_Z[(s-CD(Z+k))_+]\le\psi_k(s,C):=
 \begin{cases}
 (1-q^{A-k})(s-CB)_++\dfrac{I(s,C,u_A)}{pq^k},&k<A,\\[6pt]
 \dfrac{I(s,C,u_k)}{pq^k},&k\ge A.
 \end{cases}
\end{equation}
For each pre-reset age $z\ge\max\{A,k\}$, $W(z)$ is the mean of
$v^{m-1}$ on $[u_z,u_{z+1}]$. Its probability $pq^{z-k}$,
divided by the bin width $p^2q^z$, is $1/(pq^k)$.
Jensen therefore integrates the entire old-age tail. The younger
ages, when present, retain their common failure $B$.
No initial-age cutoff is needed.

Conditional on the first reset position $k$, the block failure is
$B^kD(Z+k)V_{r-k-1}^0$, with independent initial age and fresh
remainder. Applying \eqref{eq:paper-fresh-chord} and then
\eqref{eq:paper-jensen-pattern} gives the following bound, where
$n=r-k-1$ in the sum:
\begin{equation}\label{eq:paper-jensen-curve}
\begin{split}
 \E[(s-V_r)_+]\le\Phi_r(s):={}&q^r(s-B^r)_+\\
 &+\sum_{k=0}^{r-1}pq^k
 \left[(1-\theta_n)\psi_k(s,B^{r-1})
             +\theta_n\psi_k(s,B^kd_n)\right].
\end{split}
\end{equation}
Both comparisons preserve means, so $\Phi_r(1)=\E[1-V_r]\le r/m$.
The positive-part representation makes $\Phi_r$ convex.

\subsection{Directed bounds on the fresh comparison}

The verifier uses fixed rational bounds
$0<a_n\le d_n$, $\underline f_n\le f_n$, and
$B^n\le e_n\le1$. For $a_n<e_n$, put
\[
 \widehat\theta_n
   =\min\left\{1,\max\left\{0,
          \frac{e_n-\underline f_n}{e_n-a_n}\right\}\right\}.
\]
The chord on $[a_n,e_n]$ using the true mean has lower-endpoint
weight $(e_n-f_n)/(e_n-a_n)$. The displayed weight is at least
this large. Therefore \eqref{eq:paper-fresh-chord} remains an upper
bound after replacing its endpoints and weight by
$a_n,e_n,\widehat\theta_n$. A deterministic interval is handled
as one atom; for $n\le A$ the verifier retains the deterministic
product $B^n$ before the final scale rounding.

The recurrence for $d_n$ uses lower bounds and downward products.
The positive recurrence for $f_n$ is evaluated with outward
interval arithmetic and a lower endpoint is taken. The mixture
weight is rounded upward, with its other weight defined as exactly
the complement. Finally the two scales $B^ka_n,B^ke_n$ are rounded
downward, preserving their order, and their weights are rounded
upward. Each change is safe because $\psi_k(s,C)$ is nonincreasing
in $C$. The baseline in young atoms and the no-reset failure may
also be rounded downward; the no-reset probability is rounded
upward. Thus the fixed envelope being checked
has the form
\[
 \widehat\Phi_r(s)=\rho_0(s-c_0)_+
                   +\sum_{(k,C,w)\in\mathcal T_r}w\widehat\psi_k(s,C),
\]
with nonnegative rational weights, at most $2r$ triples, and a
baseline $\underline B\le B$ in the young atoms.
Here $\widehat\psi_k$ is \eqref{eq:paper-jensen-pattern} with
$B$ replaced by $\underline B$, while $\rho_0\ge q^r$ and
$0<c_0\le B^r$.
It is convex and dominates \eqref{eq:paper-jensen-curve}.
Outward rounding need not preserve the endpoint mean identity.

\subsection{Covering all real thresholds}

Choose $s_0>0$ such that, for every triple $(k,C,w)$,
\begin{equation}\label{eq:paper-jensen-small-cutoffs}
 s_0<c_0,\qquad s_0<C\underline B,\qquad
 s_0<C(1-u_{\max\{A,k\}}^{m-1}),
\end{equation}
where $\underline B\le B$ is the baseline failure used in young
atoms. The atoms vanish on $[0,s_0]$, and each integral uses the first
branch of \eqref{eq:paper-jensen-integral}.
Writing $C_{\min}=\min_{\mathcal T_r}C$ and using
$H_1''(t)=(m-1)^{-1}(1-t)^{-(m-2)/(m-1)}$ gives
\begin{equation}\label{eq:paper-jensen-small-bound}
 \widehat\Phi_r(s)\le
 \frac{1}{2p(m-1)}
 \left(\sum_{(k,C,w)\in\mathcal T_r}\frac{wq^{-k}}C\right)
 (1-s_0/C_{\min})^{-(m-2)/(m-1)}s^2.
\end{equation}
Since $H_r''(s)=r(m-r)^{-1}(1-s)^{(2r-m)/(m-r)}$ and its
exponent is at most $m-2$,
\[
 H_r(s)\ge\frac r{2(m-r)}(1-s_0)^{m-2}s^2
                         \qquad(0\le s\le s_0).
\]
Comparing coefficients proves the whole initial interval.

On $[s_0,1]$, take a rational partition and rational upper bounds
on $\widehat\Phi_r$ at its endpoints. Convexity bounds the function
by each joining chord, written $Ss-M$. We check that this line
lies below $H_r$ on the entire unit interval. If $S\le0$, it
suffices that $M\ge0$; if $S\ge1$, the condition is
$M-S+r/m\ge0$. For $0<S<1$, define
$u=1-S$ and $a_*=(m/r)(M-S+r/m)$.
Minimizing $H_r(s)-Ss+M$ at $s=1-u^{(m-r)/r}$ gives the exact test
\begin{equation}\label{eq:paper-prefix-powers}
                         a_*\ge0,\qquad a_*^r\ge u^m.
\end{equation}
These are rational inequalities. Every accepted chord certifies
its full interval, and explicit coverage of the partition completes
the proof of \eqref{eq:certificate-stop-loss}.

The implementation uses $128$-bit dyadic bounds for the fresh
recurrences, mixture weights, and outer probabilities; $64$-bit
downward scales; and $96$-bit endpoint upper bounds.
Arb enclosures~\cite{Johansson17} use $256$ bits.
The cutoff bounds in \eqref{eq:paper-jensen-small-cutoffs} are
rounded downward. For endpoint evaluation, an affine integral
branch is used only when its validity is certified. Otherwise,
when $s<C$, the full integral $CH_1(s/C)$ is a safe upper bound.
If $s\ge C$, the affine branch applies. Inconclusive chord signs
are resolved by exact rational powers in
\eqref{eq:paper-prefix-powers}. Endpoint computations need not
define a convex function: the chord argument uses their certified
upper bounds on the fixed convex envelope.

\section{A Finite Rational Implementation}
\label{sec:finite-implementation}

We give a finite rational implementation of Theorem~\ref{thm:main}
with the error allowances in Section~\ref{sec:parameters}. We cap and
scale the seed, replace the continuous coupling by a finite tournament,
and discretize water filling.

\subsection{Capping and scaling the seed}

Cap the age at an integer $L\ge A$, replacing $W(z)$ by
$W(\min\{z,L\})$. The capped stationary probabilities are $pq^z$
for $z<L$ and $q^L$ at the final state. The ordinary transition sends
the final state to itself on a nonreset and to zero on a reset.
The invariant used in Theorem~\ref{thm:direct-age} remains valid:
$\delta_0P^n$ now has the distribution of $\min(G,n,L)$.
All desired win probabilities decrease pointwise, so tournament feasibility
persists.

The first Bonferroni bound in the reset-failure integral gives
\[
 D(z)\le c_1q^z\quad(z\ge A),\qquad
 c_1=\frac{(m-1)p(1+q)}2.
\]

To bound the change in $\gamma$, couple the uncapped stationary age $Z$
and the capped age $\min\{Z,L\}$ using the same reset bits. This coupling
is preserved by every transition and has the correct stationary marginals.
For any coupled stationary score processes $X_k\ge X'_k$ in $[0,1]$,
product telescoping gives
\[
 0\le \E\left[\prod_{k=1}^n(1-X'_k)\right]
          -\E\left[\prod_{k=1}^n(1-X_k)\right]
       \le n\E[X_1-X'_1].
\]
The discounted number of exposures in \eqref{eq:intro-score} has mean
$m$. Hence the difference between their values of $\gamma$ is at most
$m\E[X_1-X'_1]$. This uses the stationary, unconditional score
distribution; it does not condition that distribution on previous failures.

Capping changes a score only at a reset from an age $z>L$, and the
change is at most $D(L)$. The stationary probability of such a reset is
$pq^{L+1}$. The cap loss is therefore at most
\[
                  mpq^{L+1}D(L)\le mp c_1q^{2L+1}.
\]
Scaling the capped scores by $\eta$ loses at most
$m(1-\eta)\E[X^{(L)}_1]\le1-\eta$, since capping decreases
the stationary mean from $1/m$.

For both parameter choices in \eqref{eq:paper-parameters}, take $L=5m/2$.
The rational checks in Appendix~\ref{sec:score-verifier} give
\begin{equation}\label{eq:paper-cap-bound}
 mp c_1q^{5m+1}<10^{-8}.
\end{equation}
The bounds are less than $3.047\cdot10^{-9}$ for $m=500$ and
$3.232\cdot10^{-9}$ for $m=1000$. Thus the implemented seed uses
$1251$ or $2501$ age states, respectively. For
$\eta=1-1/(m\cdot10^6)$, the scaling loss is at most
$1/(m\cdot10^6)$.
The finite matrices are rational, so their resolvent and the resulting
value of $\Gamma$ are rational as well.

\subsection{Computing the finite potential}
\label{sec:finite-potential-computation}

The inverse in \eqref{eq:algorithm-finite-potential} can be computed
by a short backward recurrence. Use the cap $C$ and scaled scores of
Section~\ref{sec:algorithm-preprocessing}, and put
\[
 a=\frac m{m+1},\qquad \rho=aq(1-b_f),\qquad
 R=\frac{1-a}{1-\rho}.
\]
Starting at the capped state, compute
\[
 \sigma_C=\frac{ap(1-W_f(C))}{1-\rho},\qquad
 \sigma_z=ap(1-W_f(z))+\rho \sigma_{z+1}\quad(z=C-1,\ldots,0).
\]
Then
\[
 \ell_0=\frac{R}{1-\sigma_0},\qquad
 \ell_z=R+\sigma_z\ell_0,\qquad h_f(z)=1-\ell_z,
 \qquad \Gamma=\sum_{z=0}^C\pi_f(z)h_f(z).
\]
Indeed, the defining equation for the miss probability is
$\ell_z=1-a+\rho\ell_{\min\{z+1,C\}}+
ap(1-W_f(z))\ell_0$. Solving its cap self-loop and substituting
backwards gives the displayed recurrence. Its denominators are
positive: $\rho<1$ and
$\sigma_0\le ap(1-b_f)/(1-\rho)<1$. Thus preprocessing uses $O(C)$
exact rational arithmetic operations for this step.

\subsection{Constructing the finite tournament tables}
\label{sec:explicit-tournament-preprocessing}

We give the full preprocessing used by $\mathrm{Tournament}$.
The input consists of $m,p,C$, the unscaled capped scores $W(z)$,
their baseline $b$, and $0<\eta<1$. All quantities below are rational.
For our parameter choices, $b>0$ and $D(C)=1-W(C)>0$.

\paragraph{Stationary block distributions.}
For each $1\le r<m$, enumerate the distribution of the scaled combined
desired win probability $Q'_r=1-\prod_{k=1}^r(1-X^f_k)$. Here $X^f_k$ is a
single-exposure score from $W_f$, starting in the stationary distribution
$\pi_f$. An exact dynamic program uses a dictionary
$\mathsf A_t(z,f)$ for the probability of age $z$ and accumulated
failure product $f$ after $t$ exposures. Initialize
$\mathsf A_0(z,1)=\pi_f(z)$. An entry of mass $a$ at $(z,f)$ sends
\[
 \begin{array}{ll}
 qa&\text{to }(\min\{z+1,C\},\,f(1-b_f)),\\
 pa&\text{to }(0,\,f(1-W_f(z))).
 \end{array}
\]
Merge equal rational keys. The failure product is a recorded value;
it does not further multiply the transition mass. After step $r$,
sum over ages with the same value $1-f$, obtaining distinct sorted
scores $\xi_{r,1}<\cdots<\xi_{r,n_r}$ and positive probabilities
$\rho_{r,j}$. One run to length $m-1$ gives all these distributions.

\paragraph{Strength bins and their means.}
Set
\[
 \delta=(1-\eta)bD(C)^{m-2},\qquad
 H=\left\lceil\frac{m-1}{\delta}\right\rceil,
 \qquad t_\ell=\frac\ell H\quad(0\le\ell\le H).
\]
For the reference strength $S_r$ with cumulative distribution $s^r$,
bin $\ell$ has probability and conditional mean
\begin{equation}\label{eq:explicit-bin-data}
 \omega_{r,\ell}=t_\ell^r-t_{\ell-1}^r,
 \qquad
 \mu_{r,\ell}=\E[S_r^{m-r}\mid S_r\text{ in bin }\ell]
       =\frac{r(t_\ell^m-t_{\ell-1}^m)}{m\omega_{r,\ell}}.
\end{equation}
All probabilities are positive and the means increase strictly with
$\ell$. Write $M_r$ for the random bin mean and $Z_r=S_r^{m-r}$.
Every power $s^{m-r}$ varies by at most $\delta$ on one bin, so
$(Z_r-\delta)_+\le M_r$ in this coupling.

Let $Q_r$ be the same capped block win before scaling. Product
telescoping gives
$Q_r-Q'_r\ge(1-\eta)bD(C)^{r-1}\ge\delta$.
Capping preserves the tournament inequalities, so
$Q_r\preceq_{\icx}Z_r$. Since $x\mapsto(x-\delta)_+$ is
increasing and convex,
\[
 Q'_r\preceq_{\icx}(Q_r-\delta)_+
       \preceq_{\icx}(Z_r-\delta)_+
       \preceq_{\icx}M_r.
\]
This supplies the finite convex-order comparison needed below.

\paragraph{The coupling linear program.}
For each $r$, the variable $K^{(r)}_{j\ell}$ is the probability of
choosing bin $\ell$ when the desired win probability is $\xi_{r,j}$. Solve
\begin{equation}\label{eq:explicit-coupling-lp}
 \begin{aligned}
 K^{(r)}_{j\ell}&\ge0,&
 \sum_{\ell=1}^H K^{(r)}_{j\ell}&=1,\\
 \sum_{j=1}^{n_r}\rho_{r,j}K^{(r)}_{j\ell}&=\omega_{r,\ell},&
 \sum_{\ell=1}^H\mu_{r,\ell}K^{(r)}_{j\ell}&\ge\xi_{r,j},\\
 \sum_{\ell=h}^H K^{(r)}_{j\ell}
       &\le\sum_{\ell=h}^H K^{(r)}_{j+1,\ell}
       &&(j<n_r,\ 2\le h\le H).
 \end{aligned}
\end{equation}
The first two lines of \eqref{eq:explicit-coupling-lp} impose
nonnegativity, the two marginals, and the
conditional-mean requirement. The last line orders the conditional
bin distributions stochastically as the desired win probability increases.
The coupling theorem in Lemma~\ref{lem:tournament} guarantees
feasibility. This bounded rational polytope has a rational feasible
point, which an exact LP algorithm can find. Store its rows as
$K_r(\xi_{r,j},\cdot)$ and check the constraints exactly.
Any coarser common rational partition with feasible versions of these
LPs also suffices; the displayed choice of $H$ guarantees feasibility.

\paragraph{The within-bin winner table.}
Conceptually, after sampling its bin, each candidate independently
draws a reference strength conditioned on that bin, and the largest
strength wins. We integrate out this continuous draw. Let $J$ be the
candidates in the highest sampled bin $[a,b]$. No other candidate
can win. For $i\in J$, its conditional winning probability is
\begin{equation}\label{eq:explicit-bin-winner}
 c_i(\mathbf r_J,[a,b])=
 \frac{r_i}{\prod_{k\in J}(b^{r_k}-a^{r_k})}
 \int_a^b s^{r_i-1}
       \prod_{k\in J\setminus\{i\}}(s^{r_k}-a^{r_k})\,ds.
\end{equation}
This is the integral of $i$'s conditional density against its
opponents' conditional cumulative distributions. The integrand is
a polynomial of degree at most $m-1$. Multiply the polynomials and
integrate each monomial exactly over the rational endpoints. The
resulting rational probabilities are nonnegative and sum to one.
Precompute them for each bin and each ordered positive multiplicity
list whose sum is at most $m$. Candidates are represented by their
positions in this list, so the tables are independent of graph labels.
List the candidates in the fixed vertex order, and map the returned
probability vector back to their labels in that same order.

Algorithm~\ref{alg:finite-tournament} now specifies the tournament
entirely with rational distributions. In particular, it does not
break equal-bin events uniformly: the multiplicities determine the
appropriate probabilities in \eqref{eq:explicit-bin-winner}.

\begin{pseudocode}{The finite calibrated tournament}{alg:finite-tournament}
\pc\>\textbf{Input:} distinct candidates $(i,r_i,Q_i)$ after all seed exposures.\\
\pc\>\textbf{if} there is one distinct candidate \textbf{then return} it.\\
\pc\>\textbf{for each} candidate $i$ \textbf{do}\\
\pc\>\>Independently draw $\ell_i$ from the table row $K_{r_i}(Q_i,\cdot)$.\\
\pc\>\textbf{end for}\\
\pc\>Set $\ell_*\gets\max_i\ell_i$ and $J\gets\{i:\ell_i=\ell_*\}$.\\
\pc\>List $J$ in the fixed vertex order and read $c(\mathbf r_J,[t_{\ell_*-1},t_{\ell_*}])$.\\
\pc\>Draw and \textbf{return} a member of $J$ with these probabilities.
\end{pseudocode}

\paragraph{Calibration and monotonicity.}
For a stationary input block, the column constraints in
\eqref{eq:explicit-coupling-lp} give the reference bin probabilities.
Lifting the bin to its original conditional strength distribution therefore
recovers the marginal $\Pr(S_r\le s)=s^r$. Opponent blocks and fresh
bin draws are independent across vertices. A candidate of strength
$s$ and multiplicity $r$ consequently wins with probability $s^{m-r}$.
Conditional only on its own desired win probability $\xi_{r,j}$, its winning
probability is at least that desired win probability by the conditional-mean
constraint. The tail constraints make its strength stochastically
increase with its desired win probability, giving the required monotonicity in
its own and its opponents' scores. The rational table integrates out
only fresh randomness, so it reproduces this kernel exactly.

The dummy uses the same seed and tables as a real vertex. A sole
candidate of multiplicity $m$ wins deterministically after consuming
all its exposures. No coupling LP is needed for that multiplicity.
The preprocessing introduces no loss beyond the single-exposure
scaling already accounted for. Its dictionary, grid, and LPs can be
very large: their sizes are not asserted to be polynomial in $m$ or
$C$, which are fixed constants in the theorem.

\subsection{Rational water filling without a known horizon}

At arrival $j$, let $W_j$ be the largest weight revealed so far and
fix an accuracy $\epsilon>0$. Every terminal deficit $d_i(x)$ is
nonincreasing and $mW_j$-Lipschitz. To see the latter, condition on
the number of exposures: each threshold credit is a Bernstein
polynomial with nondecreasing coefficients in $[0,\Gamma]$.
Its derivative is at most $m\Gamma$, and the interval of relevant
thresholds has length at most $W_j$.

Set $N_j=\lceil mj(j+1)/\epsilon\rceil$. Starting from the zero
vector, repeatedly increase by $1/N_j$ a coordinate of largest
current deficit, including the dummy, until the total probability
is one. Let $\beta_j$ be the final maximum deficit. At the last
increment of any used coordinate, its deficit is at least the then
maximum and hence at least $\beta_j$; the increment lowers it by at
most $mW_j/N_j$. It follows that
\[
 \sum_i v_id_i(v_i)
      \ge\beta_j-\frac{\epsilon W_j}{j(j+1)}.
\]
Dual feasibility still holds with $\beta_j$, and the primal--dual
budget loses only this error. Since $W_j\le\OPT$ and
$\sum_{j\ge1}1/[j(j+1)]=1$, the total competitive loss is at most
$\epsilon$. We take $\epsilon=10^{-6}$.

After $j$ arrivals, each vertex has at most $j$ threshold breakpoints.
All capped states, matrices, credits, and grid probabilities are
rational. With fixed preprocessing, the arithmetic operation count
and bit lengths are polynomial in the input size and $1/\epsilon$.

For the bit-length claim, let $D_0$ be a common denominator of
$\pi_f$ and $D_1$ a common denominator of $P_f,T_f$. Every accounting
entry after $j$ arrivals has a denominator dividing
\[
                       D_0D_1^{mj}\prod_{s=1}^jN_s^m.
\]
Splitting an interval copies its vector, and each committed mixture
update introduces at most a factor $D_1^mN_j^m$. The entries lie
in $[0,1]$, so their bit lengths are polynomial. Credit computations
add the rational input weights and fixed potential; their bit lengths
remain polynomial in the total input length.

To sample a rational probability vector, put its entries over a
common denominator $D$, draw a uniform integer in
$\{0,\ldots,D-1\}$, and use the cumulative integer masses. Draw
$\lceil\log_2D\rceil$ unbiased bits and reject values at least $D$;
the expected number of trials is less than two. The case $D=1$
is deterministic. This implements the categorical draws, reset
coins, bin draws, and winner draws using only finite bit strings.

\begin{samepage}
\section{Tightness of the Main Analysis}
\label{sec:algorithm-tightness}

This appendix concerns the ideal full-mean algorithm underlying the
competitive guarantee. Its conclusion does not assert optimality among
all online algorithms or tightness of the finite scaled implementation.

\begin{theorem}[Tightness of the ideal matching reduction]
\label{thm:ideal-triangular-tightness}
Fix $m\ge2$, $p\in(0,1)$, and an admissible nondecreasing
geometric-age map $W$ with baseline $b=W(-1)>0$ and stationary mean
$\E[X_1]=1/m$. Use exact water filling and the tournament of
Section~\ref{sec:selection}, with a kernel depending only on the current
multiset and score tuple and independent fresh randomization on each round.
Put
\[
 A_k=\E\left[\prod_{\ell=1}^k(1-X_\ell)\right],\qquad A_0=1,
 \qquad
 \gamma=1-\sum_{k\ge0}\frac{m^k}{(m+1)^{k+1}}A_k.
\]
For each $n$, let $G_n$ have offline vertices $1,\ldots,n$ and
online arrivals $1,\ldots,n$, where arrival $j$ has unit-weight
edges precisely to vertices $j,j+1,\ldots,n$. Then
\[
 \OPT(G_n)=n,\qquad
 \lim_{n\to\infty}\frac{\E[\ALG(G_n)]}n=\gamma.
\]
Consequently the worst-case competitive ratio of this ideal algorithm
is exactly $\gamma$.
\end{theorem}
\end{samepage}
\begin{proof}
The diagonal edges form a perfect matching, so $\OPT(G_n)=n$.
We first show that at arrival $j$ the candidate vector is uniform
on its $k_j=n-j+1$ neighbors. All active vertices have identical
accounting histories and, inductively, a common row $\nu$ satisfying the invariant.
Their only relevant weight interval is $(0,1]$, so their deficit
after hypothetical probability $x$ is
\[
             d(x)=\nu(I+x(T-I))^m h.
\]
This is positive: every desired win probability is at least $b$, so the geometric
stopping-time interpretation gives $h\ge mb/(1+mb)>0$, while a run of nonreset misses has positive
probability because $q(1-b)>0$. For every row $\mu$ satisfying the invariant,
\[
 \mu Th\le(1-b)\mu Ph\le(1-b)\mu h.
\]
Thus $d'(x)=m\mu(T-I)h\le-mb\mu h<0$, with
$\mu=\nu(I+x(T-I))^{m-1}$. Inactive vertices have nonpositive
deficits, as does the dummy. The unique maximizer of the strictly
concave active water-filling objective therefore assigns probability
$1/k_j$ to each active vertex and zero elsewhere. This also proves
the induction hypothesis for the next active set.

Next, singleton calibration is exact. Fix any multiset consisting of
$m$ distinct vertices, and let $g_i(x)$ be vertex $i$'s winning
probability given its singleton score $x$, averaging over independent
stationary opponent scores and fresh tournament randomness.
Calibration gives $g_i(X_i)\ge X_i$ almost surely. Since exactly one
vertex wins and every stationary singleton score has mean $1/m$,
\[
 \sum_{i=1}^m\E\bigl[g_i(X_i)-X_i\bigr]=1-m/m=0.
\]
Every summand is nonnegative, so $g_i(X_i)=X_i$ almost surely for
each vertex. This conclusion holds for every fixed distinct-vertex
multiset; symmetry of the tournament is not required.

Fix $\delta\in(0,1)$ and consider a vertex
$i\le(1-\delta)n$. Its total number of sampled exposures is
\[
 K_{i,n}=\sum_{j=1}^i\operatorname{Bin}(m,1/k_j),\qquad
 \lambda_{i,n}=\E[K_{i,n}]
      =m\sum_{k=n-i+1}^n\frac1k,
\]
where the binomials are independent. Uniformly over these vertices,
$\lambda_{i,n}$ is bounded by a constant depending only on $m,\delta$,
and $\E[K_{i,n}^2]\le\lambda_{i,n}+\lambda_{i,n}^2$.

Call the candidate tape bad for vertex $i$ if it occurs more than
once on any round, or if some opponent vertex occurs more than once
among all the rounds containing $i$. The first event has probability
at most
\[
                 \binom m2\sum_{j=1}^i k_j^{-2}.
\]
For the second, condition on all indicators specifying which draws
equal $i$, and exclude the first event. There are
$L=(m-1)K_{i,n}$ opponent draws. Each is independent and uniform on
its round's active set with $i$ removed, of size at least $\delta n-1$.
Any pair coincides with probability at most $1/(\delta n-1)$.
Hence, for $\delta n>1$,
\[
 \Pr(\text{bad for }i)
 \le\binom m2\sum_{j=1}^i k_j^{-2}
       +\frac{(m-1)^2\E[K_{i,n}^2]}{2(\delta n-1)}
       =O_{m,\delta}(n^{-1}).
\]

Condition on a candidate tape that is not bad for $i$. Its exposures
are singletons on distinct rounds, and all opponents on those rounds
are distinct vertices. The tape fixes every seed exposure index.
Each opponent contributes one stationary singleton score, and those
scores are independent because their vertex seeds are independent.
Intervening rounds do not change this statement: seed evolution
depends on sampled exposures, not on tournament winners. These scores
are also independent of the entire seed sequence of $i$.
Consequently, conditional on that sequence, exact singleton calibration
and fresh independent tournament randomness give loss probability
\[
                       \prod_{\ell=1}^{K_{i,n}}(1-X_\ell).
\]
Averaging over the target seed and then the candidate tape yields
\[
 \left|\Pr(i\text{ is never chosen})-\E[A_{K_{i,n}}]\right|
                  \le O_{m,\delta}(n^{-1}).
\]

The independent Bernoulli sum $K_{i,n}$ is within total variation
$m\sum_{j\le i}k_j^{-2}=O_{m,\delta}(n^{-1})$ of a Poisson
variable of mean $\lambda_{i,n}$. One elementary coupling proves
this bound by coupling each Bernoulli($u$) to Poisson($u$) with
error at most $u^2$, then adding independent pairs.
Uniformly for $i/n\le1-\delta$,
\[
          \lambda_{i,n}=-m\log(1-i/n)+O_{m,\delta}(n^{-1}).
\]
Since $0\le A_k\le1$, bounded Poisson expectations are continuous
in their means. Riemann summation therefore gives the limiting
contribution of these vertices as
\[
 \int_0^{1-\delta}
       \left[1-\E\bigl[A_{\operatorname{Pois}(-m\log(1-t))}\bigr]\right]dt.
\]
The remaining vertices contribute between zero and $\delta n+1$,
and the omitted integral is between zero and $\delta$. Letting
$\delta\downarrow0$ identifies the full limit. Finally, putting
$x=-\log(1-t)$ and interchanging nonnegative terms gives
\[
 \int_0^1\E\bigl[A_{\operatorname{Pois}(-m\log(1-t))}\bigr]\,dt
 =\sum_{k\ge0}A_k\int_0^\infty
            e^{-(m+1)x}\frac{(mx)^k}{k!}\,dx
 =\sum_{k\ge0}\frac{m^k}{(m+1)^{k+1}}A_k.
\]
This proves the limit. The matching reduction supplies the reverse
worst-case guarantee, so the ratio is exactly $\gamma$.
\end{proof}

\section{Reproducibility and Verification Artifacts}
\label{sec:verifier}
\label{sec:score-verifier}

The repository linked in Section~\ref{sec:introduction} contains the
numerical verifiers, generated certificates, and Lean sources.
Build instructions are in \path{README.md}; source hashes are in
\path{paper/verifiers/manifest.json}. Appendix~\ref{sec:certificate} proves the reductions
used by the tournament verifier.

The directory \path{paper/verifiers/} contains
\path{verify_score_two.py} and \path{verify_tournament_two.py}.
Run \texttt{make score} for the exact score and implementation margins,
and \texttt{make tournament} for the all-multiplicity certificate.
The latter requires Python 3.12 and the pinned dependency in
\path{requirements.txt}. Both tournament runs must report
\texttt{PASS} for every $2\le r<m$, with $12{,}915$ chords for
$m=500$ and $30{,}105$ for $m=1000$. Run the programs with Python
assertions enabled.

The formal main theorem is in
\path{research/lean/tournament1000/Main1000.lean}.
With Lean 4.19.0 and the pinned dependencies installed, run
\texttt{make lean-main} to compile its dependency closure and audit
its assumptions. The formalization has the scope and two literature
assumptions stated in Section~\ref{sec:parameters};
\path{research/lean/README.md} documents the build and its checks.

\end{document}